\def\VEL{\:{\rm km\:s^{-1}}}

\def\OIGS{\:{\rm ergs\:cm^{-2}\:s^{-1}\:\AA^{-1}}}
\def\LA{Lyman\thinspace$\alpha$}
\def\LB{Lyman\thinspace$\beta$}
\def\LG{Lyman\thinspace$\gamma$}
\def\LD{Lyman\thinspace$\delta$}

\def\OviL{\ion{O}{6} $\lambda\lambda1032,1038$}

\documentclass[12pt,preprint]{aastex}

\begin{document}

% Additional private definitions that appear to work only inside document

\newcommand{\MSOL}{\mbox{$\:M_{\sun}$}}

\newcommand{\EXPN}[2]{\mbox{$#1\times 10^{#2}$}}
\newcommand{\EXPU}[3]{\mbox{\rm $#1 \times 10^{#2} \rm\:#3$}}  % exponent with units
\newcommand{\POW}[2]{\mbox{$\rm10^{#1}\rm\:#2$}}
\newcommand{\SING}[2]{#1$\thinspace \lambda $#2}
\newcommand{\MULT}[2]{#1$\thinspace \lambda \lambda $#2}
\newcommand{\CHINU}{\mbox{$\chi_{\nu}^2$}}
\newcommand{\vsini}{\mbox{$v\:\sin{(i)}$}}
% End of defining things

\title{\bf{{\it FUSE} Spectroscopy of the White Dwarf in U Geminorum\footnote{Based on observations made with the NASA-CNES-CSA Far
Ultraviolet Spectroscopic Explorer. {\it FUSE} is operated for
NASA by the Johns Hopkins University under NASA contract
NAS5-32985} } }

\author{Knox S. Long}
\email{long@stsci.edu} \affil{Space Telescope Science Institute,
3700 San Martin Drive, Baltimore, MD 21218}

\author{Gabriel Brammer}
\email{gabriel.brammer@yale.edu} \affil{Department of Astronomy,
Yale University, \\ New Haven, CT, 06520}

\and

\author{Cynthia S.\ Froning}
\email{cfroning@casa.colorado.edu}
\affil{Center for Astrophysics and Space Astronomy, \\
University of Colorado, 593 UCB, Boulder, CO, 80309}

\begin{abstract}

Observations of U Gem with {\it FUSE} confirm that the WD is heated
by the outburst and cools during quiescence. At the end of an
outburst, the best uniform temperature WD model fits to the data
indicate a temperature of 41,000 - 47,000 K, while in
mid-quiescence, the temperature is 28,000 - 31,000 K, depending on
the gravity assumed for the WD. Photospheric abundance patterns at
the end of the outburst and in mid-quiescence show evidence of CNO
processing. Improved fits to the spectra can be obtained assuming
there is a hotter, heated portion of the WD, presumably an accretion
belt, with a temperature of 60,000 - 70,000 K occupying 14-32\% of
the surface immediately after outburst.  However, other relatively
simple models for the second component fit the data just as well and
there is no obvious signature that supports the hypothesis that the
second component arises from a separate region of the WD surface.
Hence, other physical explanations still must be considered to
explain the time evolution of the spectrum of U Gem in quiescence.
Strong orbital phase dependent absorption, most likely due to gas
above the disk, was observed during the mid-quiescence spectrum.
This material, which can be modeled in terms of gas with a
temperature of 10,000-11,000 K and a density of \POW{13}{cm^{-3}},
has a column density of \EXPU{\sim 2}{21}{cm^{-2}} at orbital phase
0.6-0.85, and is probably the same material that has been observed
to cause dips in the lightcurve at X-ray wavelengths in the past.
The discrepancy described by \cite{naylor2005_ugem} between the
radius of the WD derived on the one hand by the UV spectral analysis
and the distance to U Gem, and on the other, by the orbital elements
and the gravitational redshift remains a serious problem.
\end{abstract}


\section{Introduction \label{sec_intro}}

Dwarf novae (DNe) are binary star systems that undergo semi-regular
outbursts in which the system brightens by 3-5 visual magnitudes.
DNe consist of a white dwarf (WD) surrounded by an accretion disk of
material transferred from a low-mass late-type companion star. The
outbursts are triggered by a thermal instability in the disk that
causes an increase in the mass transfer rate
\citep{hoshi1979,mineshige1983} and can last from a day to several
weeks. In outburst, the disk is hot, ionized and optically thick and
is the dominant source of UV and optical emission. In quiescence the
disk is cool, mostly neutral, and optically thin in the continuum,
and the ultraviolet (UV) flux is usually, but not always, dominated
by emission from the WD. DNe are members of the larger class of
cataclysmic variables (CVs), all of which contain a mass-accreting
WD and whose properties are significantly affected by the magnetic
field of the WD. In DNe, the strength of the field is sufficiently
low for an accretion disk to form and extend (close) to the WD.

U Gem was the first cataclysmic variable (CV) and the first DN
discovered, and as such, is regarded as the prototypical DN. U Gem
undergoes outbursts lasting typically 7-14 days of about 5
magnitudes, reaching a peak magnitude of 9.1 about 3 times a year
\citep{szkody1984}. There are two types of outbursts, narrow and
wide, lasting $\sim$7 and $\sim$14 days, respectively
\citep{ak2002}. Unlike some prototypes, it remains a reasonable
prototype for other DNe. The WD in the system is fairly massive, 1.1
\MSOL\ \citep{sion1998,long1999}, and hot, 30,000 K in
mid-quiescence \citep{panek1984,kiplinger1991,long1993}, and does
dominate the UV spectrum in quiescence.  The UV spectrum in outburst
resembles that of a steady state accretion disk with $\dot{m}$ of
\EXPU{7}{-9}{\MSOL~yr^{-1}} \citep{panek1984,froning2001} at the
astrometrically determined distance of 100.4$\pm$3.7 pc
\citep{harrison2004}. During outburst the luminosities of the
boundary layer and the disk are similar \citep{long1996}, as
predicted by the standard theory of disk accretion in CVs
\citep{lynden-bell1974}.

{\it IUE} observations of a deepening \LA\ profile and decreasing UV
flux provided the first evidence that the WD in U Gem cools between
outbursts \citep{kiplinger1991}. However, the flux decline is less
than would occur if the entire WD cooled, and this led
\cite{long1993} to interpret the Hopkins Ultraviolet Telescope
spectrum of U Gem, the first spectrum of U Gem in quiescence that
extended to the Lyman limit, in terms of a WD with 85\% of the
surface at 30,000 K and 15\% at 57,000 K. They suggested that the
hot portion of the WD might either be due to the existence of an
accretion belt left over from the outburst that had been predicted
by \cite{kippenhahn1978}, or possibly due to an elevated accretion
rate in the disk plane following the outburst. Subsequent {\it HST}
observations have tended to confirm the observational facts. The
overall spectral shape in the wavelength range 1150-1750 \AA\ and
the character of both the \LA\ absorption profile and the depth of
the metal absorption lines seen as a result of on-going accretion
resemble that of a 38,000 K WD just after outburst cooling to 30,000
K far from outburst, but the flux evolution implies that there must
be at least two components to the spectrum \citep{long1994}.
However, the exact nature of the second component is still unclear.

In principle, observations of U Gem with {\it FUSE} can shed light
on this problem, both because the hotter component should be more
important in the {\it FUSE} spectral range 900-1187 \AA\ than in the
{\it HST} range, and because {\it FUSE} has sufficient spectral
resolution (R$\sim$12,500) to separate a slowly rotating WD from a
rapidly rotating accretion belt. Here we analyze two observations of
U Gem obtained with { \it FUSE}, the first, originally described by
\cite{froning2001}, at the end of an outburst of U Gem, and the
second during mid-quiescence, which we obtained from the {\it FUSE}
archive and has not to our knowledge been analyzed. Our primary
purpose was to better understand the processes that cause the
evolution of the spectrum of U Gem in quiescence, and especially of
the second component in the spectrum. The remainder of the paper is
organized as follows: In Sec.\ \ref{sec_obs}, we describe the
observations, our reduction of the data, and provide a qualitative
description of the spectra that were obtained.  In Sec.\
\ref{sec_analysis}, we analyze the data in terms of WD models,
explore the likely elemental abundances in the photosphere of the
WD, and try to characterize the nature of the second component in
light of complicating evidence of phase-dependent temporal
variations in the quiescent {\it FUSE} spectra.  In Sec.\
\ref{sec_discussion}, we attempt to synthesize the results in terms
of our general understanding of the UV properties of DNe in
quiescence and explore a specific discrepancy with the WD radius
inferred by different techniques. Finally, in Sec.
\ref{sec_conclusion}, we sum up.

\section{Observations and Data Reduction \label{sec_obs}}

As indicated in Fig.\ \ref{fig_aavso}, both of the observations
described here occurred when system was in optical quiescence.
However, the first observation, hereafter Obs.\ 1, was obtained just
as the system returned to quiescence, $\sim$10 days after the peak
of an outburst, whereas the second observation, hereafter Obs.\ 2,
occurred $\sim$135 days after the prior outburst peak with U Gem
well into quiescence, about 60 days before the next outburst would
occur. The two observations did not occur after the same outburst
but the nature of the outbursts was fairly similar:  Both had have
fairly ``rounded'' optical burst profiles peaking at a normal
maximum of 8.7-8.8 mag. Neither exhibited a prominent plateau. The
outburst preceding Obs.\ 1 lasted 14.9 days, based on the time above
magnitude 13, whereas the outburst before Obs.\ 2 lasted 14.1 days,
and thus both were ``wide'' outbursts. The two {{\it FUSE}}
observations were also fairly similar, with planned exposure times
of about 13,000 s, as indicated in the observational log presented
in Table \ref{table_obs}.

The {\it FUSE} spectrograph consists of four independent optical
channels that combined cover the 905-1187 \AA\ wavelength range
\citep{moos2000}. The optics of two of the four channels are
optimized for shorter wavelengths (905-1105\AA) with a SiC coating.
The optics of the other two channels are coated with LiF and
optimized for the longer wavelengths (1000-1187\AA). The data are
recorded in eight segments, A and B for each of 4 channels, and the
eight segments can be combined to cover the full 905-1187 \AA\ range
with some overlap. Both observations were taken in the
photon-counting time tag mode through the large 30" x 30" (LWRS)
aperture. This minimizes slit losses that can occur due to
misalignments of the four {\it FUSE} channels. \cite{sahnow2000}
describe the {\it FUSE} observatory and its in-flight performance in
detail.

Although Obs.\ 1 had been previously reduced by \cite{froning2001}
(denoted ``Obs.\ 4'' in that paper), we have re-reduced Obs.\ 1 and
reduced Obs.\ 2 using the {\it FUSE} data reduction pipeline
(CALFUSE 2.4.0), and combined the data from the separate channels to
produce time-averaged spectra. An important consideration in
constructing the time-averaged spectra is that {\it FUSE} guides on
a single channel, LiF1, and that thermally-induced distortions of
the optical benches can lead to significant slit losses in the
non-guided channels. To determine whether this problem affected the
U Gem data, we constructed spectra in 300 s time intervals and
compared the fluxes in the overlap regions of the various channels.
Inspection of these data showed that by the end of Obs.\ 1, U Gem
had drifted out of the apertures of all three non-guided channels.
In situations where the flux differences in the overlapping regions
is small, we rescaled the data using the strategy described by
\cite{froning2001}; if the difference was large, as it was about
half of the time for the non-guided channels, we discarded the data.
In contrast, during Obs.\ 2, none of the channels appear to have
drifted significantly and all of the data was included in the final
combined spectra. In combining the spectra, the channels were
weighted according to the errors associated with the individual
channel spectra, and regions where the flux calibration is uncertain
(i.e. the ``worm'' on LiF1B at wavelengths $>$ ~1150 \AA\
\citep{sahnow2000} was excluded).

While not seriously affecting the flux calibration, small
thermally-induced motions of the four channels, including the guided
LiF1 channel, can also induce small offsets in the wavelength
solution. Therefore in the process of combining the data, we also
checked for time-dependent errors in the wavelength solution using
narrow interstellar (IS) absorption features as fiducials.
Specifically, we measured the position of IS \ion{O}{1}
$\lambda$1039.23 and \ion{N}{1} $\lambda$1134.98 by fitting Gaussian
profiles to the observed lines. There were no obvious drifts with
time in either observation. In Obs.\ 1, the RMS variation in the
measured position of the two lines was ~0.01 \AA\ (2 $\VEL$) with a
small zero point offset of 6 $\VEL$.  In Obs.\ 2, the interstellar
lines were more difficult to measure due to a lower continuum flux
level and confusion with a (presumably) photospheric line at ~1135
\AA. Nevertheless, the RMS variation in line centers was $<$ 0.02
\AA\ (4 $\VEL$), and in this case there was no measurable zeropoint
offset.

Since most, if not all, of the FUV light from U Gem arises from the
vicinity of the WD and our primary goal is to understand the nature
of the emission on the WD photosphere, we removed the smearing
effect of the WD orbital motion by shifting the individual 300 s
segment spectra to the reference frame of the WD, using the
ephemeris of \cite{marsh1990} and $K_1 = 107.1\pm2.1$ km s$^{-1}$
and $\gamma_1 = 172\pm15$ km s$^{-1}$ obtained by \cite{long1999}
 from a series of high time resolution GHRS spectra
of U Gem (Our own analysis of the orbital parameters is discussed in
Sec.\ \ref{sec_orbit}). We shifted each 300 s spectrum using these
orbital parameters to place all of the spectra at a common velocity,
namely the recession velocity of the WD at phase=0, and we combined
the shifted spectra to obtain a time-averaged spectrum for each
observation.\footnote{Since phase 0 in the ephemeris of
\cite{marsh1990} corresponds to secondary conjunction, and since we
shifted the spectra by $\gamma_1$, the sum of the gravitational and
recessional velocity of the WD, this choice means that photospheric
lines from the WD should appear at their rest wavelengths in the
shifted spectra.} Thus, the time-averaged spectra were corrected for
the smearing of the WD spectrum due to its radial velocity motion,
while the non-moving interstellar and airglow features were smoothed
out in the process. The final time-averaged spectra were rebinned to
a wavelength resolution of 0.1 \AA\ and are shown in Fig.
\ref{fig_lines}.

As anticipated, both spectra resemble that expected from a WD with
an atmosphere that contains metals as a result of on-going
accretion.  In particular, the spectra show absorption from the H
Lyman series from \LB\ to the Lyman limit, and a rich set of metal
absorption lines.  The Obs.\ 1 spectrum peaks at 1000 \AA\ at
\EXPU{5.0}{-13}{\OIGS}. The Obs.\ 2 spectrum peaks at 1105 \AA\ at
\EXPU{2.5}{-13}{\OIGS}. The ratio of the Obs.\ 1 spectrum to the
Obs.\ 2 spectrum is greatest at wavelengths short of 1000 \AA\ (5:1
at 960 \AA) and decreases at longer wavelengths (2:1 at 1160 \AA),
indicating a reddening of the Obs.\ 2 spectrum and, thus, that the
average temperature of the WD is cooler in Obs.\ 2.
% This is
%illustrated in Fig.\ \ref{fig_shape}.
The cooling of the WD during
quiescence was noted in HUT spectra of U Gem taken 10
\citep[Astro-1;][]{long1993} and 185 \citep[Astro-2;][]{long1995}
days into quiescence. The shapes of the {\it FUSE} Obs.\ 2 and HUT
Astro-2 spectra are nearly identical considered at the ~3 \AA\
resolution of HUT. If anything, the {\it FUSE} Obs.\ 2 spectrum is
slightly redder, possibly indicating a cooler average WD
temperature. The fluxes observed on Astro-2 were slightly greater
(10\%) than observed with {\it FUSE}, also suggesting a cooler
average WD temperature at the time of {\it FUSE} observation. Given
the calibration uncertainties, however, its also possible that the
fluxes were identical in the two mid-quiescence observations. The
fluxes from the {\it FUSE} Obs.\ 1 spectrum are about 25\% higher
than observed with Astro-1 10 days into quiescence.

As shown in Fig.\ \ref{fig_lines}, absorption lines of low
ionization species of C, N, O, P, S, and Si are observed in both
observations.
%The equivalent widths of the absorption lines in both
%observations were measured using the IRAF ``splot'' routine, and the
%results, along with line identifications are shown in Table
%\ref{tab_lines}.
All of the lines seen in Obs.\ 1 appear in Obs.\ 2,
while there are additional lines of \ion{S}{2} and \ion{Fe}{3} that
appear in Obs.\ 2 and do not show up in Obs.\ 1. The lines that
appear in both observations generally have larger equivalent widths
in Obs.\ 2. Most of the lines that are seen are from ions that are
expected in the metal enriched photosphere of a WD with a
temperature of 30,000-40,000 K. The main exception is the \OviL\
doublet, which requires a temperature of at least 80,000 K, and
therefore must arise along the line of sight to the WD, but not from
the photosphere.

A comparison of the individual spectra from Obs.\ 1 shows very
little variability.  In particular, the flux measured in the line
free region between 1045 and 1055 \AA\ of the Obs.\ 1 spectrum
remains fairly constant throughout the entire integration (with an
RMS variability of  2.9\%) and no secular trends.  However, the
continuum fluxes of the 300 s Obs.\ 2 spectra vary by as much as
25\% of the time-averaged Obs.\ 2 spectrum with an RMS variation of
7.8\%. Furthermore, as indicated in Fig.\ \ref{fig_dips}, a
lightcurve of the Obs.\ 2 continuum flux plotted against orbital
phase shows dips at orbital phase 0.2-0.35 and 0.6-0.85. An Obs.\ 2
spectrum extracted around the dip between phase 0.6 and 0.85 shows a
striking increase in the number and depth of absorption features
(see the lower panel in Fig.\ \ref{fig_dips}). During the dip, the
flux below 970 \AA\ decreases substantially, and nearly all of the
lines become much more prominent. The only absorption lines that are
not noticeably stronger are those that are already quite saturated,
and \ion{O}{6}.

The coverage of Obs.\ 2 is not complete between orbital phase
0.6-0.85 and the spectra showing the increased absorption come from
a single orbit of U Gem; thus, we cannot prove that the absorption
at near phase 0.7 is orbitally dependent rather than a secular
behavior. However, dips have been seen at similar phases in soft
x-ray \citep{mason1988}, extreme ultraviolet \citep{long1996}, and
FUV \citep{froning2001} wavelengths in U Gem in outburst. More
importantly, X-ray absorption has been observed in U Gem in
quiescence near phase 0.7 by \cite{szkody1996} and by
\cite{szkody2002} using {\it ASCA} and {\it Chandra}, respectively.
 Therefore, phase-dependent absorption is the most plausible
interpretation of the time-variable absorption seen in
mid-quiescence during Obs.\ 2 in the FUV with {\it FUSE}.

\section{Analysis \label{sec_analysis}}

In order to quantify the properties of the WD in U Gem at the time
of the two {\it FUSE} observations, we have fit the spectra to a
grid of synthetic WD model spectra created using Ivan Hubeny's
TLUSTY and SYNSPEC codes for calculating the structures and spectra
of stellar atmospheres \citep{hubeny1988,hubeny1995}. The main model
grid covers a range of WD temperatures from 12,000 to 90,000 K,
gravities from log g of 8.0 to 9.0, WD rotation velocities (\vsini)
from 0 (non-rotating) to 500 $\VEL$, and metal abundances from 0.1
to 10 times the solar ratios. The synthetic spectra were computed at
fine wavelength resolution ($\delta \lambda  < $ 0.01 \AA) and
convolved with a Gaussian (FWHM = 0.1 \AA) to match the wavelength
resolution of the {\it FUSE} spectra.

Unless otherwise noted, we used a standard least-squares
minimization routine to find the models that best approximate the
data.  We assume that the reddening along the line of sight is
negligible, since that is what is expected for the value of N$_H$ of
\EXPU{2}{19}{cm^{-2}}, determined by
\cite{froning2001}.\footnote{The reddening has not been measured
directly.  \cite{verbunt1987} estimated that it is 0.0 with an upper
limit of 0.03 from the absence of a 2200\AA\ feature in {\it IUE}
spectra.} For Obs.\ 2, there is, as described earlier, time-variable
absorption. Since this extra absorption is most likely unassociated
with the WD photosphere,  we first describe fits to the portions of
the Obs.\ 2 data when this extra absorption was not present, and
return to the question of the nature of time variability in Section
\ref{sec_veil}. Here and elsewhere, when we refer to the unabsorbed
spectrum of Obs.\ 2, we mean the data outside of orbital phases
0.2-0.35 and 0.65-0.85. For Obs.\ 1, we fit the time averaged
spectrum. In fitting the data, we ignored the data near \LB\ airglow
emission and around the \ion{O}{6} lines, which are not expected in
the photosphere of a WD with a temperature characteristic of U Gem.

Based on our past experience with analyzing spectra of WDs in CVs
with {\it FUSE}, our general approach was to begin with the simplest
models, that one might reasonably expect to apply to the data, in
this case, models of uniform temperature WDs with approximately
solar photospheric abundances.  We did not really expect to obtain
good fits in a \CHINU\ sense both because the systematic errors in
the {\it FUSE} calibration exceed the statistical errors, and also
because the models themselves are subject to uncertainties in, for
example, the atomic data.  To be confident that a more complicated
model really is a better description of the actual physical
situation we required not only that \CHINU\ improve, but also that
one can point to specific regions or characteristics of the spectrum
where the more complicated model provides a qualitative improvement
in the data.  This reflects our bias that others have sometimes
relied too much on improvements in \CHINU\ alone to assert a real
physical improvement our understanding of spectrum.

\subsection{Uniform temperature WD model fits \label{sec_1T}}

\subsubsection{Solar abundance models}

We first attempted fits assuming the surface temperature of the WD
was uniform during both observations.  As a fiducial for further
fits, we first fit the spectra to models with solar abundances.
Given that we fixed N$_H$, E(B-V) and log g, the variables for the
fiducial fits were the temperature T$_{WD}$ of the WD, \vsini\ of
the WD, and N, an overall normalization.  For a WD at a distance D,
the radius R$_{WD}$ is given by D $\sqrt{N/4\pi}$.

For Obs.\ 1, as indicated in Table \ref{tab_norm}, the best fit log
g=8.5 model has a temperature T$_{WD}$ of 43,600 K, and a \vsini\ of
152 $\VEL$. Assuming that the WD is not obscured, the model
normalization combined with the known distance of 100.4$\pm$3.7 pc
\citep{harrison2004} implies the radius of the WD, R$_{WD}$ =
\EXPU{5.0\pm0.2}{8}{cm}, where here the errors are those associated
with the distance uncertainty. The model, as is shown in Fig.\
\ref{fig_obs1_1wd}, provides a reasonable qualitative, if not
statistical (\CHINU = 6.7), fit to the data. The model recreates the
shape of the continuum at wavelengths longer than ~970 \AA\
including the \LB\ and \LG\ line profiles. Many of the absorption
lines in the spectrum also have approximately the correct widths and
depths. The most noticeable failure of the model to match the
observed flux is at wavelengths short of 970 \AA\ where there is
excess flux not predicted by the model.\footnote{T$_{WD}$ is
somewhat sensitive to the reddening. Setting E(B-V) to 0.03 results
in a best fit with T$_{WD}$ of 46,400 K and \CHINU\ of 6.2; setting
it to 0.06 results in 49,700 K and \CHINU\ of 5.9.  However, the
problem of excess flux in the observed spectrum shortward of 970
\AA\ remains. The improvement in \CHINU\ with reddening has to do
with the fact that higher reddening allows a better fit to the
continuum longward of 1050 \AA} The biggest problems with features
in the spectrum are near 990 \AA, where the observed \ion{N}{3},
\ion{He}{2} and \ion{S}{3} feature is mush stronger in the data than
in the models, and near the \LB\ and \ion{O}{6} complex. As noted
earlier, a pure WD model was not expected to account for \ion{O}{6},
however the excess emission around \LB\ was not necessarily
expected. One likely possibility is that the excess is due to
emission from the disk.

% With the exception of the high-ionization NIV, SVI, and \ion{O}{6} lines,
% nearly all of the absorption features visible in the data are
% predicted by the model, though the scaled solar abundances of the
% model does not predict the correct line strengths in many cases.
% \MULT{C III}{1125, 1175} and \SING{C IV}{1169} are too strong in the
% model, while the model underestimates the strengths of the
% \SING{\ion{N}{3}}{990}, \MULT{S IV}{1063,1073}, HeII+NII $\lambda$ 1085,
% and \SING{Si III}{1113.2},, \MULT{Si IV}{1122,1128} and \SING{P
% V}{1118} are reasonably well fit by the model.  ?? Need to check the
% specific statements about what fits and what does not, since I was
% using the scaled z models for this?

% Note that the other part of the PV doublet is buried in another feature

Similar quality fits to the Obs.\ 1 spectrum can be also be obtained
with log g = 8 and log g =9 models. As enumerated in Table
\ref{tab_norm}, the best fitting log g = 8 model has T$_{WD}$ of
40,700 K, \vsini\ of 163 $\VEL$, R$_{WD}$ of \EXPU{5.3}{8}{cm}, and
\CHINU\ of 7.2. This temperature is, as one might expect, very
similar to the value of 43,410 K, obtained by \cite{froning2001}
using the same gravity and solar abundances. The differences most
likely arise from small changes in the {\it FUSE} calibration and
possibly a different selection of the exact wavelengths to fit. For
log g =9, the best fit model has T$_{WD}$ of 47,100 K, \vsini\ of
135 $\VEL$, R$_{WD}$ of 4.8, and \CHINU\ of 6.5. While all of the
model fits are unacceptable in a statistical sense, it is
interesting that the fits seem to favor higher gravities. Higher
gravity models would be expected to provide a better fit, given
estimates of the mass of the WD in U Gem.  The main reason that
higher gravity models fit the data better is that they provide a
somewhat better fit to the region of the spectrum near the Lyman
limit; other differences in a qualitative sense are quite minor. The
difference in temperature that results from the various gravities is
primarily due to changes in the profile of the Lyman lines. The
Lyman lines become more prominent as the gravity increases and less
prominent as the temperature increases. Since the spectrum one is
fitting does not change, using higher gravity models results in a
higher temperature (and a correspondingly smaller radius) for the WD
in a system.

For Obs.\ 2, the results of a similar fit to the unabsorbed portion
of the data  using solar abundance models is illustrated in Fig.\
\ref{fig_obs2_1wd}. The best-fit log g=8.5 solar-abundance model,
shown as the solid red line, has T$_{WD}$ of 30,300 K, \vsini\ of
90 $\VEL$, R$_{WD}$ of \EXPU{3.4}{8}{cm}, and \CHINU\ of 6.8.
Qualitatively, the successes and the failures of the model fit to
the Obs.\ 2 spectrum are rather similar to that of Obs.\ 1. The
model reproduces the shape of the continuum at wavelengths $>$ 970
\AA\ and the wings of \LB, but underestimates the flux at
wavelengths less than 970 \AA. The line cores of Lyman lines (\LB\
through \LD) are not well fit. All show similar profiles of excess
emission that could come from double-peaked emission from the disk.
GHRS spectra of U Gem during quiescence show evidence for
double-peaked disk emission in the core of \LA\ \citep{long1999}.
Nearly all of the metal lines present in the spectrum exist in the
model, with the exception of \MULT{S VI}{933.4, 944.5} and \MULT{O
VI}{1031.9, 1037.6} (the model line close to O VI $\lambda$ 1037.6
is \SING{C II}{1037.0}).  In contrast to the situation, in Obs.\ 1
however, it is clear the lines in the spectrum are deeper than those
in the model.  The biggest problems are with the strong absorption
features near 990\AA (which was a problem with the Obs.\ 1 spectrum
as well) and near 1085 \AA. The feature near 1085 \AA\ is primarily
due by a \ion{N}{3} triplet, as the \SING{HeII}{1085} line is much
less weaker than at the higher temperatures of the WD in Obs.\ 1.

\subsubsection{Single temperature models with scaled solar
metallicities}

We next considered the possibility that the photospheric abundance
ratios were approximately solar, but that the overall metallicity of
the photosphere was either sub- or super-solar.

For Obs.\ 1, allowing the metallicity to vary does not significantly
improve the fits, either in a qualitative or quantitative sense. The
best fits all have a metallicity of $\sim1.4$ times solar, but the
WD temperatures, radius and rotational velocities are almost
identical to those obtained when the metal abundances were solar.
Furthermore the difference in \CHINU, as a comparison of Tables
\ref{tab_norm} and \ref{tab_1wd} shows, is very small.

%In principle, this degeneracy can be broken if one assumes a
%standard WD mass-radius relation (Anderson 1988) for the WD in U
%Gem.  This was the approach used by \cite{long1999} to estimate
%the mass of U Gem based on a GHRS spectrum at mid-quiescence. They
%found \EXPU{4.7 \pm 0.6}{8}{cm} and 1.14 $\pm$ 0.07 \MSOL, or
%log g  of ??. We will return to this question later, but for now
%we will use models based for log g=8.5 as the primary set for
%comparison.

By contrast, for Obs.\ 2, allowing the metallicity to vary reduces
the value of \CHINU\ for log g = 8.5 models from 6.8 in the solar
case to 6.0 for the best fit which has a metallicity that is 3.9
times solar.  This model reproduces the strengths of \SING{S
III}{1077.1}, \MULT{S IV}{1062.7, 1073.0}, \MULT{Si III}{993.5,
1108.4, 1110.0, 1113.2}, and \MULT{Si IV}{1122.5, 1128.3} better,
and this allows a better fit to the level of the continuum as well.
As was the case for Obs.\ 1, there is a discrepancy in the fits to
the Si III line complexes. Fitting the Si III lines around 1110 \AA\
require abundances that produce lines that are too strong at 1140
\AA\ and 1155 \AA. There is evidence for a lower C abundance and
enhanced N abundance based on the fit with scaled abundances: The C
II lines $\lambda$ 1010.3 , and \SING{C III}{1175.3} are all too
strong in the scaled abundance model, and NI 1134.2, \SING{N
II}{1085.3}, and \SING{N III}{989.8} are too weak.

%The model normalization yields R$_{WD}$ = \EXPU{6.3}{8}{cm} which
%implies M$_{WD}$ = 0.93 \MSOL. [log g =8.50]. The model
%temperature is consistent with T$_{WD}$=29,900 determined by
%\citep{long1995} from HUT observations of U Gem far from outburst;
%however, the implied WD radius from those data,
%R$_{WD}$=\EXPU{5.8}{8}{cm}, is somewhat smaller than the value
%determined here [within errors if errors \EXPU{\sim0.6}{8}{cm}as
%in LG99]. ?? Need to exercise due diligence here.  I expect Long
%1995 used log g=8?? T

Based on the fits to the data there is a suggestion that the average
metallicity is higher in mid-quiescence (z = 3.3-4.7) than
immediately after outburst (z = 1.3-1.5). Whether this is a real
physical effect is unclear.  What is clear however is that there are
some elements, especially N, that have lines that are significantly
deeper than expected from our scaled-metallicity grid, and other
elements, especially C, that have lines that are weaker than
predicted from the best fits obtained from the scaled-metallicity
model grid. There is no real reason to expect that all of the metal
abundances in U Gem should scale with solar ratios, and indeed
recent analyses of {\it HST} spectra of U Gem suggest evidence for
CNO processing in the form of low C and high N abundances
\citep{sion1998,long1999}. While the 1150-1710 \AA\ spectral range
of the HST spectra contains a number of well-observed C lines, it
contains only one prominent N transition, \SING{N II}{1169.86}, and
this is in a portion of the spectrum where the {\it HST}
spectrographs tend to be less sensitive. The {\it FUSE} wavelength
range contains a different set of lines, and in particular a number
of prominent N lines. \cite{froning2001}, in discussing the spectrum
of Obs. 1, noted that the strengths of the N and C lines did
indicate that the N lines were strong relative to solar, while the C
lines appeared weaker.

\subsubsection{Abundances of individual elements}

To extend the results of the scaled metallicity modeling, we then
attempted to constrain the abundances of the elements, N, C, Si, and
S, that are responsible for most of the metal features in the
spectra. We first constructed four grids of log g = 8.5 model
spectra in which the abundances of all elements except one, either
C, N, Si, or S, were fixed at their solar values. The fitted
variables in these models therefore were $T_{WD}$, \vsini\, and the
abundance of either N, C, S, or Si. Finally, we created small grids
of models with fixed $T_{WD}$, in which all N, C, Si, and S, were
varied simultaneously. The purpose of the first part of this
procedure was determine which elements had the dominant effect on
the fits.  The purpose of the second part was to account for the
effects of having several transitions of different elements
contributing to a single feature in the spectrum.

In Obs.\ 1, the improvement in \CHINU\ in the single element fits
were most dramatic for N. In particular, for log g=8.5, changing
only the N abundance to 33 times solar reduced \CHINU\ to 5.4
compared to 6.7 for a solar or scaled abundance model. The higher N
abundance provides a much better fit to \SING{N III}{990}, though
\MULT{N III}{1002, 1003, and 1006}, which were too strong in the
solar model, are even more discrepant in the supersolar model.
Varying C alone resulted in a best fit abundance ratio of 0.2, and
reduces \CHINU\ to 6.5. The subsolar C abundance improves the fits
to the profile of \SING{C III}{ 1175} and to the strengths of the
weak \SING{C III}{1125} and \SING{C IV}{1169} lines. Allowing the S
abundance to rise to seven times solar also reduced \CHINU\ to 6.5;
this improved the fits to \MULT{S IV}{1063, 1173}.  For Si, the best fit abundance ratio was 0.8 times
solar, and \CHINU\ was 6.7, the same as for the scaled abundance fit.  One problem with Si is that the
\ion{Si}{3} $\lambda$ 1110 complex is too weak in the model even
though the \MULT{Si IV}{1122,1128} doublet is well fit.
%Since the biggest gain
%in \CHINU\ was obtained by varying N, we then created a grid of
%models with the N abundances set to 26 times solar, and refit C and
%Si. The best fit values of C and Si were about the same as
%previously, and improvements in \CHINU\ were modest.

For Obs.\ 2, like Obs.\ 1 varying the abundances of N produced a
significant reduction in \CHINU\ compared to fits with solar
abundance models. Specifically, beginning with a solar model, and
varying N, resulted in a best-fit N abundance that was 30 times
solar, and \CHINU\ of 6.2, and clear improvement over the value of
\CHINU\ of 6.8 for pure solar abundance models. As for Obs.\ 1, the
main improvement was near \SING{N III}{990}. Fitting a C abundance
of 0.2 solar decreases the \CHINU\ of the fit to 6.5, and the model
provides a much better fit to \SING{C II}{1010.3} and \SING{C
III}{1175.3}.  As was the case for Obs.\ 1, the best fit for S
abundance was high, of order 4 times solar, but the improvement in
\CHINU\ to 6.7 was modest, indicating that the spectrum is not very
sensitive to the S abundance.  Unlike Obs.\ 1, changing the Si
abundance had a large effect. Specifically, the model with a and
abundance of 4.7 times solar resulted in a \CHINU\ of 5.8. which was
not only considerably lower than for the solar case, but also less
than the value 6.0 obtained the scaled abundance models. At a
temperature of about 30,000 K, \ion{Si}{2} also contributes to the
formation of the feature at 990 \AA.

The best fits for uniform temperature WD models were obtained when
all of the abundances of all of the elements were allowed to vary
independently. For Obs.\ 1,  when a log g=8.5 30,000 K models were
generated, the best fits were obtained when C, N, Si, and S, had
abundances of 0.35, 41, 1.4, and 10 times solar; for Obs.\ 2, using
log g=8.5 43,000 models, the best fits yielded surprising similar
abundance values of 0.30, 35 4, and 6.6 times solar.  The best fit
values of \vsini\ were 150 and 250 $\VEL$ for Obs.\ 1 and 2,
respectively. The value of \CHINU\ is a shallow function of \vsini;
values of 50 $\VEL$ lower or higher produce have values of \CHINU\
that are only larger by less than 1\%.  Although general trends in
relative abundances remain the same, there is a positive correlation
of overall metallicity with \vsini. The best fits had values of
\CHINU\ of 5.0 for both Obs.\ 1 and 2, compared to the 6.7 and 6.0
for the models with scaled metallicities.  The best fits are shown
as the black lines in Figures \ref{fig_obs1_1wd} and
\ref{fig_obs2_1wd} for Obs.\ 1 and 2. The improvements in the model
fits are generally localized to the lines, as one would expect, and
the overall improvement in \CHINU\ is quite significant, but not
enough to provide a good statistical fit to the data.

Our basic conclusions with regard to abundances are that the {\it
FUSE} spectra do provide strong support for CNO processing of
material in U Gem, consistent with previous modeling efforts
\citep{sion1998, long1999,froning2001}, and a strong suggestion of
Si overabundance. The spectra also hint at S overabundance as well,
and the apparent overabundance is large, but the identifiable
affects on the spectra are small, and hence we feel this result to
be fairly uncertain.  As shown in Figures \ref{fig_obs1_1wd} and
\ref{fig_obs2_1wd}, it is also important to point out that there are
examples where some transitions of the an ion are well modeled, but
others are not, suggesting either additional components to the
absorption, or limitations in the synthetic spectra.

Regardless of which set of abundances are utilized in fitting the
{\it FUSE} spectra of U Gem, the uniform temperature model fits to
the Obs.\ 1 and Obs.\ 2. spectra indicate that the WD has cooled by
12,000 - 16,000 K from the end of the outburst to mid-quiescence,
depending on the value assumed for log g. This drop in temperature
is greater than the more typical value of 8,000 K that has been
reported previously analyses with other UV spectrographs, e.g. {\it
HST} \citep{long1994}, and HUT \citep{long1995}.  The apparent
radius of the WD is about 15\% larger in Obs.\ 2 than Obs.\ 1; this
result is consistent with the previous studies, and was in fact
first seen with {\it IUE} \citep{kiplinger1991}; it is one of
several reasons for considering more complicated models for the UV
spectra of U Gem, especially in the immediate post-outburst period.

\subsection{Two-component WD fits\label{sec_2T}}

As noted above, uniform T modeling of the spectrum of U Gem just
after outburst and far from outburst suggest a larger radius for the
WD far from outburst than at outburst.  This is essentially a
restatement of fact originally commented on by \cite{kiplinger1991}
that the WD flux is falling more slowly than suggested by the
apparent change in the temperature of the WD, and which led
\cite{long1993} to suggest that a hot accretion belt might exist on
the WD.

Using the HUT, Long et al.\ (1993,1995) found that
single-temperature WD models underestimated U Gem's UV flux below
970 \AA, similar to the failure of the WD models described in
Section \ref{sec_1T} to accurately predict the flux at the short
wavelength end of the {\it FUSE} spectra. The discrepancy in the
HUT analysis was mitigated by adding a second high temperature WD
component to the model that covered 15\% of the WD surface close
to outburst and 1\% of the surface far from outburst.

Consequently, we carried out fits of to the data from Obs.\ 1 and 2
assuming two separate regions on the WD surface.  We allowed
different metallicities and different rotational velocities in each
region of the white dwarf surface and carried out fits for log
g=8.0, 8.5, and 9.0. The results are summarized in Table
\ref{tab_2wd}. The results for log g=8.5 are typical. In this case,
allowing two WD components in the fit to the Obs.\ 1 spectrum, we
find a cool component with T$_{WD}$=28,500 K that covers 82\% of the
WD surface and a hot component with T$_{WD}$=70,000 K that covers
18\%. The cool and hot model components have scaled abundances of
1.5 and 8.9 times solar and WD rotation rates of 87 and 243 $\VEL$,
respectively. \CHINU\ improves to 5.7 from 6.2 in the corresponding
single component model. The improvement in \CHINU\ is primarily due
to an improvement in the fit at the shortest wavelengths. The higher
temperature component dominates the flux throughout, as indicated in
Fig. \ref{fig_obs1_2wd}, but especially at the shortest wavelengths.
A WD photosphere with a temperature of 60,000 to 70,000 K has fewer
lines than one with lower temperature and so the two temperature
fits generally favor a more metal enriched atmosphere than one with
solar abundances.  The lines are fairly well fit with the two-T
model, although \ion{N}{3} 989 and \ion{C}{3} 977 remain a problem.
In this particular fit, the rotational velocity of the higher
temperature components is somewhat higher, 243 $\VEL$, than the
lower temperature component, 87 $\VEL$ as expected if the hot
component is rapidly rotating.  But this is clearly not a robust
result, since the log g = 9 fit of the same type has the cooler
component rotating more rapidly.

A two component fit to the Obs.\ 2 spectrum, using the log g=8.5
model grid also yields a modest improvement \CHINU, 5.7 compared to
6.0 for the case of a single component with variable abundances. The
lower temperature component has T$_{WD}$=26,100 K, and an abundance
that is 5.7 times solar. It covers 81\% of the WD surface, very
similar to the percentage covered by the cool component in Obs.\ 1.
The higher temperature component has a temperature of 34,500 K.  The
total normalization for the WD with this fit is \EXPN{5.7}{-23},
which corresponds to a radius of \EXPU{6.6}{8}{cm}, compared to
\EXPN{5.5}{-23} and \EXPU{6.5}{8}{cm} for a similar fit to the Obs.\
1 spectrum. In the case of the best fit model for log g=8.5, the two
temperature fit seems to resolve the problem with a WD that grows in
radius during quiescence.  A comparison of the fits obtained for log
g = 8 and log g = 9 yields a similar results. Thus, the 2-T WD model
fits to the {\it FUSE}{ data do seem to provide some modest support
for the idea that there is a heated region on the surface of the WD.

The argument that a second source in the spectrum of U Gem arises
from the WD surface would be stronger if it could be shown that a
competing model gave a less significant result.  One alternative
would be residual disk emission, but unfortunately our understanding
of how to model an accretion disk in or near quiescence is very
primitive.  Therefore, we opted to see whether a simple power law
model for the second component would produce a better or worse fit
to the data.. The variables for this fit were $T_{WD}$, z, \vsini,
and the normalization of the WD, plus a power law index and
normalization for the second component. The best fits for Obs.\ 1
and 2, assuming log g=8.5 for the WD  had \CHINU\ of 5.5 and 5.6
respectively, just slightly worse than for the 2-T WD model fits.
The implied WD temperatures were similar, 41,000 K and 29,200 K, to
those obtained for the single T WD models. Qualitatively, as shown
in Fig.\ \ref{fig_obs1_wd_pow}, the model fits looked rather similar
to those obtained for the 2-T WD model fits.

On the basis of this analysis, we conclude that while there are real
departures in the shape of the spectrum from a simple LTE WD model
in U Gem immediately after an outburst, a physical interpretation in
terms of two-temperatures on the WD surface is not demanded by the
data.  We will return to the nature of the second component in Sec.\
\ref{sec_wd_cool}.

\subsection{Phase-dependent absorption \label{sec_veil}}

The phase-resolved spectra of Obs.\ 2 show clear evidence of
variable line absorption.  As illustrated in Fig.\
\ref{fig_abs_1125}, the same lines contribute to the absorption
during all phase intervals.  Furthermore, many of the same lines
appear in the portions of the phase-resolved spectra we have
designated as ``unabsorbed''. The phase 0.6-0.85 absorption is very
similar to that between phase intervals 0.2-0.35, except that the
lines are deeper in the phase 0.6-0.85 spectrum.  The lines are not
exactly at zero velocity with respect to the WD. At phase 0.2-0.35,
the lines are blue-shifted (in the frame of the WD) by $\sim 50
\VEL$ while at phase 0.6-0.85 the lines are red-shifted by $\sim 120
\VEL$.  These wavelength shifts are important since they imply that
the absorption is not the result of changes in the photosphere
itself.

In an attempt to characterize the absorption, we have modeled the
Obs.\ 2 spectra in terms of a WD photosphere and a ``slab'' or veil
of overlying material.  For simplicity, we have assumed solar
abundances and LTE conditions in the slab material. Neither of these
conditions is likely to be correct in detail, but alternatives are
all more complicated and without some physical model seem
unjustified at this time. Each slab is described by its density,
temperature, turbulent velocity and column density. Our procedure
for modeling the absorption of the slab is as follows. Using an
option of Ivan Hubeny's SYNSPEC program, we first calculate
opacities as a function of wavelength in the slab as a function of
density and temperature. To account for the effects of turbulence,
we then smooth the opacities, and calculate the transmission of the
slab as a function of wavelength. We also shifted the spectra by
either -0.18 or 0.44 \AA, to account for the offset of the
absorption lines in the observed spectra. We created a grid of
models for temperatures from 5,000 to 25,000 K, for densities ($N_H
= N_{HI}+N_{HII}$) ranging from \POW{9}{cm^{-3}} to
\POW{13}{cm^{-3}}, for turbulent velocities $v_{turb}$ ranging 0 to
300 $\VEL$, and for column densities ranging from log N$_H$ of 18 to
23. In attempting to fit the data, we assumed that the underlying
continuum was generated from the WD photosphere and that the
photosphere had solar abundances.

In attempting to the fit the data, we initially used a standard
$\CHINU$ minimization technique and fit the same portions of the
data that we had used in the previous fits.  However, this resulted
in fits that fell well below the observed spectrum where there is
little or no absorption, especially in the phase 0.6-0.85 spectrum.
The reason this occurs is that a standard $\CHINU$ fit heavily
weights the points with the smallest errors, which are the data
points with greatest absorption, dragging the model continuum down
in instances where the model is unable to reproduce all of the
absorption lines.  Therefore we opted for an approach that we
believe gives a better ``eye-ball'' description of the data at the
expense of formal statistical correctness. Specifically, we have
adopted a two-pass approach to fitting the data, which consists of
using an initial standard $\CHINU$ minimization fit to screen out
highly discrepant points, namely points with an initial $\CHINU$ of
25 or greater. We then refit the remaining data points (about 95\%
of those considered in the initial fit) to the models to find a fit
that describes most of the data points. This results in fits that
follow the shape of the continuum well and approximates most, but
not all, of the lines. There are a number of other ways that can be
used to obtain fits that qualitatively represent the data including
limiting the contribution to total $\CHINU$ rather than eliminating
discrepant points, or using an asymmetric metric that gives extra
weight to data points where the model underestimates the data. These
techniques produce similar results qualitatively, although they tend
to yield best fits with somewhat larger line widths, expressed as
turbulent velocities, in the fits described below. Our impression is
that the lines widths using our preferred technique are a more
accurate representation of the data.

We applied this technique to each of the Obs.\ 2 spectra, the
``unabsorbed'' spectrum, the phase 0.20-0.35 spectrum, and the phase
0.60-0.85 spectrum.  Results of the fits assuming normal abundance
log  g = 8.5 WD atmospheres and slabs with densities of
\POW{13}{cm^{-3}} are shown in Fig.\ \ref{fig_abs_fit} and tabulated
in Table \ref{tab_veil}.  The WD temperatures, 30,500 K for the
``unabsorbed spectrum'', 29,400 K for the phase 0.20-0.35 spectrum,
and 29,900 K for the phase 0.6-0.85 spectrum are close to the value
of 29,700 K derived for a simple uniform temperature WD model with
variable abundances.  For densities of \POW{13}{cm^{-3}}, the
effective temperature of the veil was about 10,000-11,000 K for both
the phase 0.2-0.35 and the phase 0.6-0.85 spectrum. As expected, the
column density of ionized and unionized hydrogen was higher in the
fit to the phase 0.6-0.85 spectrum (log N$_H$ = 21.3) than in the
phase 0.20-0.35 spectrum (20.7). To first order the properties of
the slab are the same during both periods when absorption is
observed. Similar results, both in terms of the qualitative nature
of the fits and in terms of the column densities are obtained when
other slab densities are considered. Specifically, the derived WD
temperatures are similar and the column densities derived for the
slab are similar. However, the temperature derived from the plasma
is somewhat higher, 12,000-13000 K for a density of
\POW{9}{cm^{-3}}, instead of 10,000-11,000 K.  The higher
temperature that is required with lower densities is a direct
consequence of the simplifying assumption that the gas is in LTE.

A disturbing possibility that must be considered is that the
absorption is not confined to phase 0.2-0.35 and 0.6-0.85. This is
hard to rule out completely, but Fig.\ \ref{fig_abs_fit} does
provide a certain amount of comfort.  The effects of the slab on the
``unabsorbed'' spectrum are relatively minor compared to those seen
in the fits to the phase 0.20-0.35 and 0.60-0.85 spectra.

\subsection{Orbital parameters from the {\it FUSE} data \label{sec_orbit}}

The orbital parameters of the WD in U Gem have been measured several
times.  The most detailed study was carried out by \cite{long1999},
who used the GHRS to obtain a series of time-resolved GHRS spectra
covering the wavelength range 1168-1448 \AA\ to derive a value of
K$_1$ of $107\pm2.1 \: \VEL$. \cite{long1999} found that low
ionization-state lines of \ion{C}{2}, \ion{Si}{2}, and \ion{Si}{3}
had an average $\gamma$ velocity of 172.1$\pm$15 $\VEL$, whereas the
higher ionization-state lines of \ion{Si}{4} and \ion{N}{5} had
lower values of 124$\pm$15 $\VEL$ and 102$\pm$10 $\VEL$,
respectively. Since the low ionization stale lines are expected in
the photosphere of a 30,000 K WD, they concluded that these lines
provided a good measurement of $\gamma_1$, that is the velocity
shift due both to the recessional velocity of the U Gem system and
the gravitational redshift of the WD surface.  They suggested that
the higher ionization state lines were formed at a location above
the WD surface. \ion{N}{5} is not expected in a WD photosphere with
$T_{WD}$ of 30,000-40,000 K, and the \ion{Si}{4} lines were stronger
than predicted for the Si abundance derived from \ion{Si}{2} and
\ion{Si}{3}. The results of their study were in agreement with the
result reported slightly earlier by \cite{sion1998} based on
observations of the \ion{Si}{3} multiplet at 1300 \AA\ at two
specific phases in the orbital period; Sion et al.\ found K$_1$ and
$\gamma_1$ to be 107 $\VEL$ and 161 $\VEL$, respectively, but gave
no error estimates. As previously noted, we used this K$_1$ velocity
to produce the average spectra for spectral analysis.

In principle, the {\it FUSE} observations described here provide an
independent measurement of K$_1$, since they have good phase
coverage and since {\it FUSE} has more than sufficient resolution to
measure velocities in this range, and so we attempted such an
analysis. Here, we used unshifted 300 second spectra.\footnote{For
this portion of the analysis, we used spectra created with CALFUSE
3.1.}. We restricted our analysis to data obtained with the LiF1
channel since this was the channel used for guiding. We rebinned the
original data to 0.1~\AA\ to improve the S/N somewhat. We measured
the central wavelengths of several of the strongest absorption lines
in each of the individual time-resolved spectra from Obs.\ 1 and
Obs.\ 2.  For this we used the IRAF SPECFIT task described by
\cite{kriss1994}. We fit the lines in SPECFIT using Gaussian line
profiles and taking into account the errors in the rebinned spectra.
For Obs.\ 1, we fit the \ion{Si}{4} $\lambda$1122~\AA\ and
\ion{Si}{4} $+$ \ion{P}{5} $\lambda$1128~\AA\ transitions.  For
Obs.\ 2, we fit these transitions and added the \ion{Si}{3}
$\lambda$1110 and 1113~\AA\ lines (we also fit the 1108 line but the
fits were poor due to low S/N and were not used).  We compared the
central wavelengths of each line to the wavelength center in the
time-averaged LiF1 spectrum for each observation.

We then converted the wavelength shifts to velocities and fit a sine
function with the appropriate period to all the lines in an
observation to determine the radial velocity amplitude K$_1$.  We
allowed the amplitude and phase of the sine curve to vary, but not
the period (or eccentricity). After the initial fit, we created a
time-averaged spectrum with the orbital motion removed, recalculated
the central wavelengths of each line, found new velocity shifts for
the lines and refit the wavelength shifts. This cycle was repeated
several times until the fits central wavelengths in the
time-averaged spectra were stable.\footnote{One could have
alternatively fit the line centers to a amplitude, a phase, and a
velocity offset.  This technique avoids the iterative process that
we describe here, and indeed yields similar results for K$_1$.
However, the technique we used yielded better \CHINU\ than a
non-iterative fit to a single amplitude, offset, and phase for all
of the lines, presumably due to the fact that the measurements of
the line centroids of the average spectrum were measured more
consistently as a result of our iterative approach.}
Figure~\ref{fig_k1} shows the best fits for Obs.\ 1 and Obs.\ 2. The
uncertainty on the amplitude represents the range of amplitudes that
yielded fits within $\chi^{2}+$4.61 of the best fit to establish
90\%, or 1.6 $\sigma$, confidence limits \citep{lampton1976}.
Finally, for Obs.\ 2, we repeated the fits using only the unabsorbed
phases: 0 -- 0.2, 0.35 -- 0.6, and 0.85 -- 1.

For Obs.\ 1, the best fit sine curve has an amplitude K$_1$ of
122$\pm$10~km~s$^{-1}$ and a phase offset from the ephemeris of
\cite{marsh1990} of  0.04$\pm$0.01.  The fit had $\chi^{2}_{\nu}$ =
1.2. For Obs.\ 2, the best fit sine curve has a larger amplitude of
132$\pm6$~km~s$^{-1}$ and a phase offset from the ephemeris of
-0.04$\pm$0.01, with $\chi^{2}_{\nu}$ = 2.4, but this is most likely
affected by the effects of the additional absorption discussed in
Section \ref{sec_veil}. If the fit is restricted to data from
``unabsorbed'' phases, then the amplitude drops to
117$^{+9}_{-8}$~km~s$^{-1}$ , which is close to that obtained for
Obs.\ 1, even though  $\chi^{2}_{\nu}$ remains quite high at 2.3.
The quality of the fits is shown in Fig. \ref{fig_k1}.  Alternative
approaches to obtaining K$_1$, such as simple cross-correlation
measurements gave very similar results. Thus the {\it FUSE} data
suggest a slightly higher value of $K_1$ than the two {\it
HST}-based studies. However, the errors on the {\it FUSE} $K_1$
velocities are fairly large (as a result of the fact that {\it FUSE}
is a much smaller telescope than {\it HST}), and the {\it HST} and
{\it FUSE} values differ formally at less than 2$\sigma$.

Next, we attempted to calculate absolute wavelengths for lines by
measuring their central wavelengths in a time-averaged spectrum with
the orbital motion removed, using the amplitudes calculated above.
First, we determined corrections to the absolute wavelength solution
by measuring the central wavelengths of several interstellar lines
(\ion{O}{1} $\lambda$1039~\AA, \ion{Ar}{1} $\lambda$1048~\AA, and
the \ion{N}{1} $\lambda$1134~\AA\ triplet) in the original
time-averaged spectrum.  The absolute wavelength offset corrections
were small:  4~km~s$^{-1}$ for Obs.\ 1 and 8.5~km~s$^{-1}$ for Obs.\
2. These offsets are very similar to the values we had obtained in
our original reduction of the data with CALFUSE 2.4, discussed in
Sec.\ \ref{sec_obs} . (Note that the reduced {\it FUSE} spectra are
already corrected for a heliocentric motion as part of the
calibration pipeline). We then measured the central wavelengths in
the U~Gem lines in the orbital motion-corrected spectrum and
compared them to their laboratory values.  To focus on WD motion
rather than that of any intervening material, we used only the
non-absorbed spectra and adopted 118~km~s$^{-1}$ as the K$_1$
amplitude in Obs.\ 2.

We measured \ion{S}{4} $\lambda$1062~\AA, \ion{S}{4}
$\lambda$1073~\AA, \ion{Si}{4} $\lambda$1066~\AA, \ion{Si}{3}
$\lambda$1108~\AA, \ion{Si}{3} $\lambda$1110~\AA, \ion{Si}{3}
$\lambda$1113~\AA, and \ion{Si}{4}$\lambda$1122~\AA\ for both
observations.  We  omitted the $\lambda$1128~\AA\ transition because
it is a blend of \ion{Si}{4} and \ion{P}{5}. We initially measured
the wavelengths of each transition assuming Gaussian profiles for
each of the lines.  All of the transitions did show positive
$\gamma$ velocities measured in this manner. However, the $\gamma$
velocities range from about 65 $\VEL$ for \ion{S}{5} $\lambda$1062
to a maximum of 155 $\VEL$ for the \ion{Si}{3} $\lambda$1108. And it
was immediately clear that this approach led to significant
inconsistencies in the $\gamma$ velocities of individual components
of the same multiplet, especially \ion{Si}{3}.
% Results of these measurements are shown in
%Table \ref{tab_gamma}and the portions of the spectra that were
%measured are shown in Fig.\ \ref{fig_gamma_vel}. The error bars in
%the Table here represent uncertainties in the measures of the line
%centers in SPECFIT only and are larger in Obs.\ 2 because the source
%was fainter. The results of this analysis were not wholly
%satisfactory. All of the transitions are redshifted, as had been
%found previously, but the shifts are not consistent for individual
%multiplets. Some of the reasons for this are obvious from a detailed
%inspection of Fig.\ \ref{fig_gamma_vel}
Several of the lines are obviously asymmetric, and this clearly
explains why for example, the shift for \ion{S}{4}
$\lambda$1062~\AA, when measured form a Gaussian fit to that
feature, was less than the other two members of that multiplet in
Obs.\ 1, probably as a result of a contribution to this line from
another line. And in Obs.\ 2, its clear that one is affected by the
effects of absorption as the lines are broader and sometimes appear
to have multiple minima.

Therefore, in the end, we elected to measure the minimum flux value
of each of the transitions.  Results of these measurements are shown
in Table \ref{tab_gamma} and the portions of the spectra that were
measured are shown in Fig.\ \ref{fig_gamma_vel}.  For Obs.\ 1, the
average value of $\gamma$ is 144.9 $\VEL$ and the standard deviation
from the mean is 13.9 $\VEL$; for Obs.\ 2, the average is 131.2
$\VEL$ and the standard deviation is 10.2 $\VEL$. These values are
close to the values of $\gamma$ reported by \cite{long1999} and by
\cite{sion1998} using GHRS, but they do not show the pronounced
change in $\gamma$ velocity with ionization state reported by them.
None of the ionization lines have $\gamma$ velocities as great as
measured by them for \ion{Si}{3}, which they argue corresponds to
$\gamma_1$ of the WD photosphere. As was the case for the
measurement of K$_1$, the difference is however significant at most
at the 2 $\sigma$ level.

A possible way to bring the measurements into closer agreement would
be to question the absolute wavelength scale. The {\it FUSE}
observations were made through the LWRS aperture, and so in
principle, the absolute wavelength scale can be in error by as much
as 0.25 A, or about 65 $\VEL$. However, the typical error is thought
to be less than this. Bowen (2005, as quoted on the {\it FUSE}
website) has compared velocities of H$_2$ lines measured with {\it
FUSE} to interstellar \ion{Cl}{1} $\lambda$1347 and finds a mean
error of $+$10$\pm$6 $\VEL$. We have attempted to compensate for
offsets in the wavelength scale by referencing our wavelength scale
to those of IS lines. Nevertheless, this could be a problem.
\cite{long1999} note that the core of Ly$\alpha$, which they presume
is IS, has a $\gamma$ velocity of 39$_{-30}^{+10} \: \VEL$. The core
of Ly$\alpha$ is IS in origin. The \ion{N}{1} lines are also IS.
Assuming the velocity shifts of all of the IS lines are the same, we
would need to add 30 $\VEL$ to our velocities to put them on the
GHRS wavelength scale.  If that is the case, then our mean $\gamma$
velocities would be much closer to those derived with GHRS for low
ionization state lines.

In view of these uncertainties, our conclusion is that the orbital
parameters derived from the analysis of the {\it FUSE} data are not
to be preferred to {\it HST} values, even though they do suggest
that if a capability for high resolution UV spectroscopy is restored
to {\it HST}, that it would be desirable to remeasure especially the
$\gamma$ velocity of the WD.  As will be discussed in Section
\ref{sec_mass}, there is currently a discrepancy between the radius
of the WD derived from the normalization of the spectrum  and the
radius implied by the gravitational redshift, and the latter
requires an accurate measure of $\gamma_1$.

\section{Discussion \label{sec_discussion}}

\subsection{WD Cooling\label{sec_wd_cool}}

The {\it FUSE} observations confirm once again
\citep{kiplinger1991,long1993,long1994} that the WD in U Gem cools,
or appears to cool, between outbursts. The cooling is apparent in
the decline in FUV flux, the fact that the flux at short wavelengths
declines more than at long wavelengths and the fact that \LB\ is
broader far from outburst. The average cycle time for outbursts of U
Gem is 132 days \citep{ak2002}.  The only detailed study of a single
interoutburst interval was conducted with {\it IUE}
\citep{kiplinger1991} and that study appears to show that the (1620
\AA) UV flux declines slowly (with some scatter) throughout the
entire interval. Unless STIS is recommissioned or COS installed on
{\it HST} on an upcoming Shuttle mission, it seems unlikely that
this situation will change.  This is unfortunate since it makes
separating the physical process that contribute to the flux decline
difficult.

Cooling of the WD is observed in other systems.  The best examples
of this are probably, VW Hyi and WZ Sge.  In VW Hyi, the WD is
heated to either 23,000 K in a normal outburst or 27,000 K in a
superoutburst.  It then cools back 19,000 K with an exponential
decay time constant of 2.8 or 9.8 days for a normal or
superoutburst, respectively \citep{1996gaensicke_vwhyi}.   The
differences in the two situations are presumably associated with the
fact that superoutbursts deposit more and more matter on the WD, and
last longer than normal outbursts. In this regard, typical outbursts
of U Gem including the outburst that preceded Obs.\ 1 are more like
superoutbursts of VW Hyi in terms of integrated energy and duration.
Since typical outbursts in VW Hyi are separated by 28 days
\citep{ak2002}, we cannot follow long term cooling trends in VW Hyi.
WZ Sge represents the opposite extreme. It went into outburst in
2001, the first time in 22 years. The WD was heated to 26,000 K (at
least), and has in the past 4 years cooled with a time constant of
about 180 days to 15,000 K
\citep{2004long_wzsge_hst,2006godon_wzsge_hst}, close to its
pre-outburst temperature of 14,800 K \citep{1997cheng_wzsge_ghrs}.
The WZ Sge outburst lasted about 24 days (followed by a series of
echo outbursts); this and the very long interoutburst period
presumably account for the long decay time constant.

A variety of processes are likely to contribute to the heating and
cooling of the WD, and disentangling these processes is one of the
main challenges of CV research today.  The mechanism that seems most
likely to dominate on long time scales (and a process that allows
the creation of detailed models) is compression heating; this is the
physical response of the WD to the deposition of additional mass on
the WD surface
\citep{sion1995a,2002townsley_compression_heating,godon2002,piro2005_wd_heating}.
The WD is hotter than before, due both to the release of
gravitational energy as the star rearranges its internal structure
and to slow burning of material at the base of the accreted
envelope. \cite{sion1995a} showed in particular that the basic
properties of the WD in U Gem, heating by of order 10,000 K and
cooling that had timescales of months, could be produced for
plausible accretion scenarios. Other processes that could also be
involved include direct heating of the outer atmosphere of the WD
during the outburst \citep{pringle1988} and elevated accretion just
after an outburst, perhaps associated with a coronal flow
\citep{1994meyer_coronal_flow}.  Direct heating during the outburst
affects the outermost layers of a WD and is expected to be a
short-term phenomenon and as a result is not expected to be
important in U Gem even a week or two after the outburst.  However,
\cite{2006godon_wzsge_hst} have had difficulty in explaining the
slow decline in the temperature of the WD in WZ Sge without ongoing
heating of the WD via continued accretion.

Of the well-studied systems, U Gem is unique in that UV flux does
not decline as rapidly as expected if the emission arises solely
from a uniform temperature WD with fixed radius. This is apparent in
the {\it FUSE} analysis and had been seen previously in HUT and {\it
HST} spectra \citep{long1993,long1994}. By contrast, similar
analyses of WZ Sge show that all of the post-outburst spectra are
consistent with a fixed radius \cite{2004long_wzsge_hst}. Since it
seems unlikely on physical grounds that the radius of the WD in U
Gem is actually growing during quiescence, alternative explanations
are needed. There are four basic escapes from this dilemma: (a) to
argue that the temperature of the WD is not uniform, (b) to argue
that the WD is partially obscured during the first observation, (c)
to argue that there is a separate source that causes the problem,
and (d) to argue that the discrepancy is not sufficiently large to
worry about at this time.

The main advantage of solutions to the time-variable radius problem
that involve the a non-uniform surface temperature distribution is
that it explains why the spectrum qualitatively resembles that
expected from WD. The main theoretical challenge of this kind of
interpretation is how to credibly create and maintain the asymmetry
on the WD surface once the dwarf nova outburst is over, especially
since the readjustment of the internal structure of the WD is
basically a spherically symmetric process (at least for a slowly
rotating WD). \cite{long1993} suggested two ways to maintain a
hotter region of the WD surface: preferential heating of the portion
of the WD surface in a boundary layer near the disk plane powered by
ongoing accretion and slow release of kinetic energy stored in a
rotating accretion belt spun-up during the preceding outburst.

At that time, the importance of compression heating was not
recognized as it is today, so \cite{long1993} assumed that
difference in luminosity just after outburst and in mid-quiescence
had to be fully explained. Since the extra luminosity was
\EXPU{3}{32}{ergs~s^{-1}}, this implied an accretion rate of
\EXPU{1.7}{15}{g~s^{-1}}, far greater than would have been derived
from the X-ray luminosity of \EXPU{1.1}{31}{erg~s^{-1}}
\citep{szkody1996}.  They were also concerned that if the accretion
rate were this high, then observational signatures of the disk
should have been seen in the HUT (850-1850 \AA) spectra. Today, it
is less clear that these specific problems rule out continued
accretion as the cause of the distortions in the spectrum of U Gem.
However, most of the evidence today is that accretion on the WD is
fairly spherical. In particular, while in outburst the boundary
layer is thought to be optically thick and geometrically
thin,\footnote{Geometrically thin may be a misnomer for recent
calculations by \cite{fisker2005_boundary_layer} suggest that the
boundary layer in outburst expands to cover a significant fraction
of the WD.} in quiescence the boundary layer is expected to be
optically thin and geometrically thick. High resolution X-ray
observations of U Gem \citep{szkody2002} and other systems show that
the X-ray emission arises from material that is not rotating with
the inner disk, which suggests that accretion of this gas, if it
occurs at all, is close to spherical.

The basic problem with the accretion belt hypothesis is that there
has been little or no detailed modeling of this phenomenon since the
pioneering work of \cite{kippenhahn1978} and
\cite{kutter1989_accretion_belt}, and (to our knowledge) no modeling
of the specific effects resulting from time variable accretion seen
in a dwarf nova outburst.\footnote{\cite{piro2004_spreading} have
recently discussed a spreading layer that could moves hot recently
accreted material from the equator toward the pole. This could cover
up to about 10\% of the WD surface at the peak of an outburst and
could merge into an accretion belt. But the timescale for this
spreading layer to remain a distinct entity is quite short, and as
they note, unlikely to account for result a multi-temperature WD
surface even ten days after the peak of an outburst.} The idea of an
accretion belt, which was posited to explain aspects of nova
explosion, is that the viscosity of WD envelope is low and therefore
that material arriving at the WD surface with Keplerian velocities
will spin up the outer layers of the WD near the disk plane. The
size and extent of the rotating region would be determined by an
instability at the interface between the rotating hydrogen-rich
accreting material and outer layers of the WD, which would result in
an accretion belt extending, according to \cite{kippenhahn1978},
about $\pm$20\degr\ from the disk plane.  It is not clear how
important accretion belts are in the context of nova explosions
\cite[see, e.g.][for a recent discussion]{porter1998_rotatingWD}. In
any event, \cite{long1993} suggested that the kinetic energy
released in this process might be what was observed in U Gem just
after outburst. They pointed out that the ``smoking gun'' for this
explanation would be the detection of lines, particularly higher
ionization state lines, that clearly showed evidence of rapid
rotation.

There have been a number of attempts to model quiescent systems
other than U Gem in terms of the WD and a more rapidly rotating
second component.  For example, \cite{2001sion_vwhyi} analyzed {\it
HST} spectra of VW Hyi in quiescence and showed that \CHINU\ was
improved if the spectra were modeled in terms of a two component WD
(or a WD and a rapidly rotating inner disk annulus), better in terms
of \CHINU\ than in terms of a uniform temperature WD.  More
recently, \cite{2004godon_vwhhyi_fuse} analyzed a quiescent spectrum
of VW Hyi obtained with {\it FUSE}, in terms of a two-component WD.
They found that the spectrum could be fit in terms of one component
with a temperature of 23,000 K rotating with \vsini\ of 400 $\VEL$
and s second component with a temperature of 50,000 K rotating at
3000 $\VEL$. This led \cite{2005godon_vwhyi_belt} to suggest that an
accretion belt had been detected in VW Hyi.  But a careful
examination of the spectra described in both of the cases above
shows that while there is a clear improvement in \CHINU, the
improvements result from small changes in profile shapes of a large
set of lines as well as  the overall shape of the spectrum. There is
no example of which we are aware in which an individual feature that
shows rapid rotation is identified, and as a result, we are not
convinced that there is evidence for a rapid rotation in a second
component to the emission. A totally featureless second component
would likely have produced a similar improvement in \CHINU. This
applies to the {\it FUSE} observations of U Gem also. Furthermore,
we see no clear trends in the widths of individual lines with
ionization potential or with observation. This is borne out by the
fits as well.  One might have hoped, as a result of the higher
spectral resolution of {\it FUSE} (R$\sim$12,500) as compared to HUT
(300) or {\it HST} ($\sim$ 1200 for the U Gem observations), that a
rapidly rotating second component might be more apparent. But the
best two-component fits do not consistently indicate that the second
higher temperature component, if it exists, is rotating more rapidly
than the lower temperature portion of the WD surface.

It should be noted at this stage that it would be possible to create
a non-rotating belt if most of the light that is observed from the
second component to the emission is reradiated.
\cite{fisker2005_boundary_layer} have carried out simulations of the
boundary layer in non-magnetic CVs, and suggest that there might be
a slowly decaying source of emission from the boundary layer just
after outburst during the transition to quiescence. They seem to
have in mind a source that is directly observable. They also suggest
this is a source might fully account for the long-term cooling of
the WD, but as we noted if compression heating is operative, this is
not necessary. Nevertheless, if there is a hot boundary layer, it is
possible that the second component that we do see is light created
in the boundary layer that  is re-radiated from slowly rotating WD
surface.

The second possibility is that the WD photosphere has a uniform
surface temperature, but that the WD is partially obscured by the
disk just after outburst, but not far from outburst.
\cite{1994meyer_coronal_flow} have suggested that the inner disk
extends close to the WD surface immediately after outburst, but that
the inner disk evaporates in the early portion of the quiescent
period. In this context it might be possible to explain a growth in
the apparent radius of the WD. Ignoring limb darkening, the
fractional reduction in flux from a WD with the ``bottom'' half of
the WD obscured is given by $1/2 + 1/2 \; cos(i)$ where $i$ is the
inclination. For an inclination of 65\degr\ the flux would be
reduced to 71\% that of an unobscured WD, and the implied radius
would be 84\% of the true radius of the WD.  The difference in radii
in Obs.\ 1 and Obs. 2 are of order 10\%, and therefore obscuration
could explain the apparent growth in WD radius.

Despite the fact that the order of magnitude estimate above
indicates that obscuration by the disk if it extended to the
interior could obscure the lower portion of the WD, we are skeptical
that this is the explanation.  The portion of the disk that would
have to occult the WD would be located within 1 WD radius of WD
surface, and this region is the illuminated (if by nothing else) by
the full radiation field of an approximately 40,000 K blackbody.

The third possibility in our list is that there is a second source
that causes the WD temperature estimate just after outburst to be
too high. This could come about if there is a second component which
distorts the spectrum at the shortest wavelengths or if the WD
models we (and others) have used are simply not adequate to model
the spectrum. The WD in U Gem is not that of a normal WD. The matter
is being continually accreted on the surface and in the case of
Obs.\ 1., the face has recently been buffeted by the outburst that
preceded it.  If the temperature were lower than we have estimated,
then the normalization would have to increase to match the observed
flux.  If we were observing the Rayleigh-Jeans tail of the WD
spectrum, then the normalization would scale inversely as the
temperature; to increase the normalization by 10\% would require a
temperature decrease of 10\%.  Model fits in which the normalization
was constrained to be \EXPU{4.5~(5.5)}{-23}{sr} imply a T$_{WD})$ of
44,000 (40,000) K instead of the value 46,700 K obtained for Obs.\
1., when the normalization is not constrained.

Given the uncertainty in the models, the fact that \CHINU\ is not
close to one for any type of model we explored, and the possibility
of a distorting second component in the spectrum, we do not feel the
either a change in WD radius or a multi-component temperature on the
WD surface is demanded by the data.  This is essentially the last in
our list of four possibilities.  What is needed at this stage is a
better set of data with a number of observations taken after a
single outburst.

\subsection{Phase-dependent Absorption
\label{sec_phase_dependent_absoprtion_discussion}}

The mid-quiescence spectrum of U Gem shows time-variable absorption.
The absorption is greatest near phase 0.7, but is also observed near
phase 0.2. Phase-dependent absorption had been reported in the FUV
previously in outburst spectra by \cite{froning2001}, but this
represents the first time such absorption has been observed in the
FUV in quiescence.  As is the case of the mid-quiescence spectrum,
variability in the outburst spectra was due to changes in relatively
narrow (250-800 $\VEL$ FWHM) lines from ions such as Si III and S
IV. In outburst, the FUV spectra of U Gem and other DNe are
dominated by emission from the rapidly rotating inner disk and
therefore \cite{froning2001} argued that the material producing the
absorption had to be elevated above the photosphere of the outer
disk. Although line depths were largest between phase 0.53 and 0.79,
absorption was observed throughout three separate observations of a
single outburst.  This implies that the absorbing material is not
confined to a single azimuthal region of the disk. If this were the
case, in quiescence it would certainly be a concern for abundance
analysis assuming lines were formed in the photosphere.

Phase-dependent absorption in U Gem has also been observed in
X-rays, both in outburst and in quiescence. In their study of U Gem
in outburst with {\it EXOSAT} \cite{mason1988} fitted changes in the
flux near phase 0.7 in various energy bands as additional absorption
due to cold material, equivalent to N$_H$ of \EXPU{3}{20}{cm^{-2}}.
However, the {\it EUVE} observations analyzed by \cite{long1996}
indicate that the continuum source is almost fully obscured and the
emission that remains are photons scattered by a wind that extends
above the surface of a disk which appears thicker at some orbital
phases than others.  In quiescence, observing with {\it ASCA},
\cite{szkody1996} saw a 50\% drop in the 0.5-2 keV X-ray flux near
phase 0.7.  The absorption was far less at higher energies, and
\cite{szkody1996} concluded that the data were consistent with an
X-ray source of order the size of the WD, and extra absorption
equivalent to N$_H$ of \EXPU{3.6}{21}{cm^{-2}} at phase 0.7.  This
is roughly consistent with the value of \EXPU{2}{21}{cm^{-2}} that
we infer from our analysis in Sec.\ \ref{sec_analysis}.

Phase-dependent absorption in CVs is generally understood to be a
consequence of the interaction between the disk and the stream of
material from the secondary star. This is also the explanation for a
similar phenomenon in a class of compact low-mass X-ray binaries,
known as ``X-ray dippers'', which also show absorption near orbital
phase 0.7. \cite{lubow1976} were the first to discuss the
possibility that gas flowing over the disk from the secondary would
have a vertical scale height substantially larger than the standard
scale height of the disk. \cite{frank1987_dipping}, in the context
of X-ray binaries, were the first to suggest that thickening of the
disk near the circularization radius ($\sim$\POW{10}{cm}) rather
than at the edge of the disk, and to predict that ``dips'' rather
than full occultations of the central X-ray source should occur in
the inclination range 60-75\degr. In the case of U Gem, Doppler
images clearly show a stream penetrating well inside the outer edge
of the disk with velocities intermediate between those expected for
an unimpeded stream and co-rotation with the disk \citep{marsh1990}.
\cite{hirose1991_dips} carried out the first 3-D particle
simulations of disks, indicating that the ratio of the vertical
height of the disk was 10-20\% of the disk radius and is greatest
near orbital phases 0.8 and (to a lesser degree) 0.2, which is what
we see in the Obs.\ 2 {\it FUSE} data. More recently,
\cite{kunze2001MNRAS_dipping} have carried out SPH-simulations of
stream overflow, including cases with the system parameters of U
Gem, indicating that a substantial fraction of the material settles
at 30-40\% of the distance from the WD to inner Lagrange point, an
indicating that material can reach altitudes of 20-25\degr\ of the
disk plane.  No one, to our knowledge, has reported  the line of
sight velocities of the material along the line of sight to the WD.
This would be quite interesting, since the {\it FUSE} data shows
absorption lines that are redshifted by about 120 $\VEL$ at between
phases 0.6-0.85, and blues shifted by about 50 $\VEL$ at phases
0.2-0.35 with respect to the WD.

\subsection{CNO-processed material in the WD photosphere}

Despite the time variable-absorption that was observed in the Obs.\
2 spectra, the fact that fits to both the Obs.\ 1 and the unabsorbed
portion of the Obs.\ 2 were improved using models with large N
overabundances and sub-solar C abundances provides strong support
for the existence of CNO processed material in the WD photosphere of
U Gem The confirmation using {\it FUSE} data of earlier suggestions
arising primarily from {\it HST} data \citep{sion1998,long1999}is
important because there is only one strong N line in the {\it HST}
wavelength range, \SING{NIII}{1184}.

U Gem was one of the first CVs for which a large N overabundance was
suggested based on an analysis of abundances on the surface of the
WD, but there is increasing evidence that a significant fraction of
CVs exhibit anomalous abundance ratios, and more specifically large
N overabundances \cite[see, e. g.][]{gaensicke2003}.  Evidence for
CNO processed material has been reported not only from UV spectra of
the WDs in CVs, but also in IR spectroscopy of some CV secondaries
\citep[including U Gem,][]{harrison2005_secondary}, and in UV
spectra of the disks of some in the form of anomalously large NV:CIV
line ratios \citep{Mauche1997,gaensicke2003}. This and the fact that
heavy elements quickly sink below the WD photosphere
\citep{paquette_1997_diffusion} suggests that the CNO material on
the WD surfaces of CVs is accreted from the secondary. Two sources
of this material have been proposed: (a) a secondary that was
originally massive and survived the thermal mass transfer stage,
possibly leading to a supersoft X-ray stage \citep{schenker2002},
which is now bringing CNO-enriched material to the surface from the
core by convection, and (b) nova-explosions that pollute the
atmosphere of the secondary \citep{marks1997}.  At present, it
unclear which of these suggestions is correct. \cite{sion2001_vwhyi}
did report the discovery of large overabundances of P and an a
general abundance pattern in one {\it HST} spectrum of VW Hyi that
suggest material from the thermonuclear runaway expected in a nova
explosion, but this has not (to our knowledge) been seen in any
other system. Furthermore, even if it is correct in this case it is
not clear that it could account for the bulk of the systems in which
CNO-processed material has been observed.

\subsection{Radius and Mass of the WD in U Gem\label{sec_mass}}

\cite{long1999} used 1162-1448 \AA\ {\it HST}/GHRS spectra to
determine a radius of \EXPU{4.7\pm0.6}{8}{cm} and inferred from this
a WD mass of $1.14\pm0.07$ \MSOL.  They based their determination on
log g=8.5 model estimate of the normalization factor of
\EXPN{4.11}{-23} and a distance of 82$\pm$13 pc derived from the
Bailey's (1981) method. Using the mid-quescence {\it FUSE} spectrum,
single temperature, scaled-abundance models,j and a distance of
100.4$\pm$3.7 pc \citep{harrison2004}, we find a radius of
\EXPU{5.7_{-0.2}^{+0.5}}{8}{cm}.  Assuming the WD in U Gem obeys a
standard mass-radius relationship \citep{anderson1988} and that the
surface of the WD is fully visible, the {\it FUSE} analysis leads
directly to a mass estimate of $1.00^{+0.04}_{-0.05} \MSOL$, where
the error bars here are determined simply by the results of the
various gravities in the models.\footnote{Using Obs.\ 1, the radius
is \EXPU{5.1_{-0.3}^{+0.2}}{8}{cm} and the mass is
1.10$_{-0.04}^{+0.02}$ \MSOL. But the \cite{long1999} measurement
was made in mid-quiescence, as was the case for Obs.\ 2.} The
results are not consistent with one another. Why? The answer is
solely that the distance has increased by 22\%. \cite{long1999} used
the \cite{bailey1981} relation to establish the distance of
82$\pm$13 pc for an inclination of 67\degr, whereas we have used the
new astrometric distance, which should be more reliable. With the
larger distance, the radius derived by \cite{long1999} would have
been \EXPU{5.7\pm0.8}{8}{cm}, almost identical to the values
obtained with the {\it FUSE} data. This is not surprising since the
measured T$_{WD}$, observed fluxes, and indeed the models used to
analyses the data are similar.

\cite{naylor2005_ugem} have recently conducted a detailed study of
the secondary star in U Gem.  They find $K_2$ to be 300 $\VEL$, very
precisely, in agreement with earlier values 309$\pm$3 $\VEL$
\citep{friend1990ugem} and 283$\pm$15 $\VEL$ \citep{wade1981ugem}.
They also find an accurate value of 29$\pm$6 $\VEL$ for $\gamma_2$,
somewhat lower than the value of 46$\pm$6 $\VEL$, obtained by
\cite{friend1990ugem}, and considerably lower than the value of
84.9$\pm$ 9.9 $\VEL$ obtained by \cite{wade1981ugem}. From the value
of $\gamma_1$ for the WD obtained by \cite{long1999} from the lower
ionization-state lines in the GHRS spectra of U Gem, they derive a
gravitational redshift $\gamma_{grav}$ of 143$\pm$15 $\VEL$. Based
on this determination of for $\gamma_{grav}$ for the WD,
\cite{naylor2005_ugem} concluded that R$_{WD}$ was, reading directly
from Fig.\ 8 of that paper, \EXPU{3.9\pm0.4}{8}{cm}, if the WD in U
Gem obeys the Hamada-Salpeter mass radius relationship
\citep{hamada1961}, or alternatively, \EXPU{3.7\pm0.9}{8}{cm} if the
inclination of U Gem between 62 and 74\degr. The conundrum uncovered
by this analysis is the photospheric radius derived from the fits to
{\it HST} spectra of U Gem corrected to reflect the astrometric
distance is about 50\% larger than the gravitational radius. The
photometric radius \EXPU{5.7_{-0.2}^{+0.5}}{8}{cm} we derive from
{\it FUSE} spectroscopy, does not change this picture.

The basic situation is shown in Fig.\ \ref{fig_massrad}, which is
similar to Fig.\ 8 of \cite{naylor2005_ugem}, based on the {\it
FUSE} results described here. (The slightly higher value of K$_1$
obtained with {\it FUSE} implies a somewhat higher WD mass for a
fixed inclination.) For specificity, suppose the actual inclination
is 67\degr.  Then M$_{WD}$ is 1.26\MSOL\ and R$_{WD}$, based on
$\gamma_{grav}$, should be \EXPU{3.8}{8}{cm}, whereas R$_{WD}$
inferred from the {\it FUSE} spectral analysis of Obs.\ 2 is
\EXPU{5.7_{-0.2}^{+0.5}}{8}{cm}, or 1.5 times larger.  This means,
since the flux scales with R$_{WD}^2$, that the observed flux is 2.3
times larger than expected.  This is a large difference.  To obtain
the observed flux, T$_{WD}$ for Obs.\ 2 would have to be increased
to $\sim$38,000 K, assuming all other aspects of our analysis are
correct.   The shapes of a the 38,000 K model spectrum is
qualitatively different from that observed with {\it FUSE} in Obs.\
2. It seems very unlikely that this can explain why the radius
derived from the spectral analysis is so much larger than predicted
from the orbital parameters and $\gamma_{grav}$.

Photometric determinations of the radius are crucially dependent on
the estimate of distance. Indeed Fig.\ 7 of \cite{long1999}, which
is very similar to our Fig.\ \ref{fig_massrad} contains no hint of a
difficulty reconciling the photometric radius with that predicted by
the \cite{hamada1961} relationship. It is interesting in this regard
that \cite{schreiber2002_sscyg} have had difficulty in explaining
the fact that SS Cyg does not show standstills in view of the larger
mass transfer rate implied by an upward revision of the distance to
SS Cyg based on \cite{harrison2004}. It is possible that the
astrometric distance derived by \cite{harrison2004} is incorrect,
although that certainly seems unlikely. In any event,it would be
useful in this regard to have an independent parallax distance for U
Gem (and SS Cyg).

Flux-based determinations of R$_{WD}$ also depend upon the
assumption that the synthetic spectra used to compare with data have
not only the correct shape, but also the correct surface fluxes.
This is an assumption that could be questioned in view of the fact
that we generated spectra from very simple pure H, LTE atmospheres.
We have, however, performed several tests to assure ourselves that
the surfaces fluxes are not significantly affected by the simplicity
of our assumption about the structure of the atmosphere. First, we
performed tests in which we compered spectra from atmospheres
calculated assuming solar abundances to those calculated from H. At
the temperatures and gravities appropriate for U Gem, the fractional
differences in the surface fluxes were quite small, less than 5\% in
the {\it FUSE} wavelength range. Second, we created a model spectrum
for Sirius B using the T$_{WD}$ and gravity obtained by
\cite{barstow2005_siriusb}.  We normalized the spectrum using
parallax distance and R$_{WD}$ for Sirius B.  Our model fluxes are
within about 15\% of the fluxes observed with {\it HST} in the
wavelength range 1780-1930 \AA. Third, to check whether different
sets of opacities or a different code might produces a significantly
different flux, we compared spectra generated using Tlusty/Synspec
with our simple LTE assumptions with \cite{kurucz1992} model
spectra. At temperatures near 30,000 K, we found good agreement
between the Kurucz model spectra (which are admittedly for lower
gravities) and those generated with TLUSTY/SYNSPEC. In particular,
at wavelengths between 1050 and 1700 \AA\ (selected to cover the
spectral ranges analyzed here and by Long \& Gilliland), the spectra
agree in terms of overall normalization to an accuracy of about
20\%. Hence, insofar as we can determine, the disagreement between
the photometric radius and the gravitational radius is not due to
inadequacies in the model spectra being used for the analysis.

If the solution is not in the WD models or the distance, and if the
determination of $\gamma_{grav}$ is correct, then one is left to
argue that there is some second source in U Gem. The most obvious
possibility is the disk,  and, as we have noted, there are some
evidence of emission from the disk, in double peaked excesses of
emission at the position of the Lyman lines. In SS Cyg and WX Hyi
where continuum emission from the disk is seen, emission is
accompanied by broad emission lines from resonance lines of
\ion{N}{5}, \ion{Si}{4}, and \ion{C}{4}
\citep{lonb2005_sscyg_wxhyi}. There is no evidence of this in {\it
HST} spectra of U Gem in quiescence.   Aside from the excesses near
the Lyman lines, there is no evidence for a rapidly rotating
component in U Gem. If the emission arises from the inner disk, the
lines widths would be very broad and it is hard to reconcile this
with the deep, relatively narrow absorbtion lines observed in the
{\it FUSE} spectra.   The second component would also have to have a
spectrum that mimicked that of a WD.  In VW Hyi where the FUSE
observations show two distinct emission components the second
component is most visible at the shortest wavelengths and it does
not show the deep Lyman line profiles of a WD
\citep{2004godon_vwhhyi_fuse}.

One can of course question the measurements of $\gamma_{grav}$, but
to bring the photometric radius and gravitational radius into
agreement, one would one would need to reduce $\gamma_{grav}$, to
about 100 $\VEL$. Since $\gamma_{grav} = \gamma_{1} - \gamma_{2}$,
one can question the determinations of either of $\gamma_{1}$ of the
WD or $\gamma_{2}$ the secondary, or both. \cite{naylor2005_ugem}
used their value of $\gamma_2$ of 29$\pm$6 $\VEL$, which is the
lowest of all of the determinations of $\gamma_2$ in conjunction
with the highest value of $\gamma_1$ of 172.1$\pm$ 15 $\VEL$
determined by \cite{long1999} from low ionization state lines, and
the discrepancy of between the photometric radius and the
gravitational radius is therefore maximized by the choice. The
$\gamma_{1}$ of the WD that is derived from the {\it FUSE} data is
lower than that obtained by \cite{long1999} and by \cite{sion1998},
but the difference is not nearly enough. Furthermore, even in the
absence of the measurement of $\gamma_{grav}$, the photospheric
radius appears to be significantly larger than the Hamada-Salpeter
mass-radius relationship suggests. (The inclination of U Gem cannot
be greater than 74\degr or it would fully eclipse.) The
Hamada-Salpeter relation is calculated for cold WDs, and for lower
masses the effects of finite temperature significantly alter the
expected radii, but for higher masses this effect is small.  At 1.1
\MSOL\ for example, \cite{wood1995} finds that a 100,000 K C/O core
WD is only about 10\% larger than a cold WD of the same mass, not
enough to account for the radius obtained from the flux, the
effective temperature, and the distance.

Therefore, we, like \cite{naylor2005_ugem}, do not have a good way
to explain away this problem.  It is quite possible that a number of
factors contribute, which suggests that a number of the observation
needs to be repeated.

\section{Summary \label{sec_conclusion}}

In this study, we have reanalyzed {\it FUSE} spectra of U Gem
obtained by \cite{froning2001} at the end of an outburst and
performed the first analysis of spectra obtained in mid-quiescence
after a different but similar outburst.  Our primary goal was to
contrast the two sets of spectra in order to learn more about the
response of the WD to the outburst.  The principle surprise in the
analysis of the mid-quiescence spectra was the discovery of large
phase-dependent absorption in the spectra, which complicated the
analysis of the WD spectra, but which provides additional
information about material observed previously at X-ray wavelengths
that must be located at large distances from the disk plane. Our
main conclusions are as follows:
\begin{itemize}
\item Both the post-outburst and mid-quiescence spectra are
dominated by the WD, as had been apparent from earlier observations
with HUT and {\it HST}.  The WD, when the {\it FUSE} spectra are
analyzed in terms of a uniform temperature WD, appears to cool from
41000-47,000 K, depending on gravity, to about 30,000 K. with higher
gravities suggesting higher temperatures.  Although
multi-temperature WD fits improve the fits in a \CHINU\ sense, the
data do not require multiple temperatures, especially since none of
the fits that we have carried out result in values of \CHINU\
approaching a value of 1.  There are a variety of alternatives,
additional components to the spectrum as well as inadequacies in the
WD models, that could explain the apparent deficiencies in the model
fits.

\item We find a K$_1$ velocity for the WD of approximately 120 $\VEL$.
This is close to but slightly larger than the values deduced by
\cite{long1999} and by \cite{sion1998}.  The difference in K$_1$
only minimally affects the determination of WD mass. There appear to
be some differences in the $\gamma$ velocity derived from different
lines, as suggested by the analysis of \cite{long1999}, and our
measurement of the $\gamma$ velocity is somewhat smaller than
determined by \cite{long1999} or \cite{sion1998}.

\item  The abundance analyses of both the post-outburst and the
mid-quiescent spectrum confirm CNO enrichment of the material being
accreted onto the WD photosphere.  The actual values of the
abundances, especially the large overabundances that are found in
some cases, are probably suspect, but the basic conclusion is not.
In particular, it is clear that one must be concerned about the
problem of absorption material within the system.

\item The absorbing material that is seen preferentially near phases 0.2 and  0.7 in
the mid-quiescence spectrum is due to ionized material with an
effective temperature of 10,000-11000 K if the density of the gas is
about \POW{13}{cm^{-3}}, a few thousand degrees hotter if the
density is \POW{9}{cm^{-3}}. The same material is probably also
responsible for the absorption seen in X-rays previously.

\item Our analysis of the  {\it FUSE} data  reenforces the fact that
there is a severe and unexplained discrepancy between the
photometric radius derived from the FUV flux observed in
mid-quiescence, the temperature derived from spectral fits to the
spectrum, and the astrometric distance obtained by
\cite{harrison2004}, and the radius inferred from mass determined
from K$_1$, K$_2$, and either $\gamma_{grav}$ or the Hamada-Salpeter
mass-radius relationship.
\end{itemize}

\acknowledgments{This analysis of {\it FUSE} data would not have
been possible without the financial support from NASA through grants
NAG5-9283 and NNG04GQ38G. We appreciate this, as well the dedicated
efforts of the entire {\it FUSE} team. We also appreciate the
efforts of an anonymous referee who made numerous constructive
comments on the original manuscript.}

%\appendix
%\pagebreak
%\include{tab1}

\pagebreak

%figure1
\figcaption[aavso]{Optical lightcurves of U Gem at the time of the
{\it FUSE} observations using data assembled by the AAVSO (E. O.
Waagen 2004, private communication). \label{fig_aavso}}

%figure 2
%\figcaption[shape]{The time-averaged {\it FUSE } spectra of U Gem
%obtained during Obs.\ 1, at the end of an outburst, and Obs.\ 2, in
%mid-quiescence.  The vertical axes of the individual panels are
%scaled so that the long-wavelength portion of the spectra are in
%similar physical locations in the figure in order to emphasize the
%differences in the shape of the spectra. \label{fig_shape}}

%figure 3
\figcaption[lines]{The {\it FUSE} spectra of U Gem obtained during
Obs.\ 1 (black) shortly after outburst and Obs.\ 2 (grey) when the
system was far from outburst with the lines labeled. Most of the
lines observed in Obs.\ 1 also appear in Obs.\ 2, including some
high ionization lines, e.g. \OviL. \label{fig_lines}}

%figure 4
\figcaption[dips]{ The upper panel shows the normalized flux in the
wavelength range 1045-1055 \AA\ from U Gem in Obs.\ 2  as a function
of orbital phase. The phase intervals that were used to construct
the ``unabsorbed'' spectrum are shown in black. The lower panel
shows the ``unabsorbed'' spectrum in black and the spectrum obtained
during phase intervals 0.6 to 0.85 in grey.   The 0.1 A spectra were
smoothed with a 0.5 A boxcar to create this figure.
below.\label{fig_dips}}

%figure 6
\figcaption[obs1_wd1]{Uniform temperature solar abundance WD model
fits to the Obs.\ 1 spectrum.  The data are plotted in grey; the
best log g  = 8.5 model, with abundances that were individual varied
is plotted in blue. For comparison, the best fit model assuming log
g = 8.5 and solar abundances is plotted in red The regions of the
data that were excluded from the fitting are shown in a lighter
shade of grey. The positions of various lines are indicated. Lines
that are clearly seen in the spectrum are indicated by thicker
labels. The narrow emission feature centered on Lyman$\beta$ is due to airglow.\label{fig_obs1_1wd}}

% figure 7
\figcaption[obs2]{Uniform temperature WD model fit to the Obs.\ 2
spectrum.  The format of the figure is identical to that of Fig.\
\ref{fig_obs1_1wd}. The best fit model assuming log g = 8.5 and
solar abundances is plotted in red. The best fit when individual
abundances of  was allowed to vary is plotted in
blue. The narrow emission feeatures at Lyman$\beta$ and Lyman$\gamma$ are due to airglow. \label{fig_obs2_1wd} }

%figure 8
\figcaption[obs1_2wd]{A two-T log g=8.5 fit to the Obs.\ 1 spectrum
of U Gem. The data and the model fit are shown as in the previous
figures.  In addition, the contribution of the low and higher
temperature components are shown as the dotted and dashed lines,
respectively. The higher temperature component dominates throughout,
but especially at the shortest wavelengths. \label{fig_obs1_2wd}}

\figcaption[obs1_wd_pow]{Similar to Fig.\ \ref{fig_obs1_2wd}, except
for a model consisting of a WD and a powerlaw second component.
\label{fig_obs1_wd_pow}}

%\figcaption[obs2_2wd]{Comparison of single-component
%T$_{WD}$=30,410 K (solid line) and two-component T$_{WD}$=27,500
%and 47,000 K (dashed line) model fits to Obs.\ 2.
%\label{fig_obs2_2wd}}

%\figcaption[obs1-obs2]{Obs.\ 1 minus Obs.\ 2 difference spectrum,
%fit with a 73,600 K WD  model. \label{fig_obs1-obs2_2wd}}

%figure 8
\figcaption[abs_1125]{A comparison of spectra obtained during Obs.\
2 that we have called ``unabsorbed'' (black) compared to the that
observed during Phase 0.2-0.35 (red) and Phase 0.6-0.85 (blue).
\label{fig_abs_1125}}

%figure 9
\figcaption[abs_fit]{Attempts to fit the spectra during the
``absorbed'' phase of Obs.\ 2 with an LTE slab absorbing the light
of the WD. The upper, middle, and lower panels show fits to the
``Unabsorbed'', phase 0.2-0.35, and phase 0.6-0.85 spectra,
respectively.  The data are plotted in grey and regions that were
excluded in a lighter grey.  The solid red line corresponds to the
fit obtained assuming that a slab of material lies along the line of
sight to a WD with log g  of 8.5 and solar abundances. The solid
green line is the WD with the absorption removed.  Although only a
100 \AA\ region from 1078-1178 \AA\ region is shown, the fit was
based on the entire spectrum. \label{fig_abs_fit}}

\figcaption[k1.ps]{Radial velocity fits to the FUV spectra of U Gem.
The upper panel shows the velocity shifts of the Obs.\ 1 spectrum
vs.\ orbital phase with the best fit radial velocity curve with an
amplitude of 122$\pm$10 $\VEL$.  The lower panel shows the velocity
shifts of the Obs.\ 2 spectrum and the best fit radial velocity
curve, with an amplitude of 132$\pm$6 $\VEL$; if the portions of the
data obtained during phases 0.2-0.35 and 0.6-0.85 are excluded the
best fit value of K$_1$ in Obs.\ 2 drops to 117$^{+9}_{-8}$ $\VEL$.
In both panels, the velocity shifts of the \protect\ion{Si}{4}
1122~\AA\ line are plotted with filled circles and the velocity
shifts of the \protect\ion{Si}{4} 1128~\AA\ line with open
triangles.  In the lower plot, the shifts of the \protect\ion{Si}{3}
$\lambda$1110 and 1113~\AA\ lines are plotted with open squares and
open circles, respectively. \label{fig_k1}}

% fig 11
\figcaption[fig_vshift.eps]{Selected regions of the spectra obtained
from the LiF1 channel for Obs.\ 1 and 2.  The spectra plotted in
black have orbital motion, here assumed to be 120 $\VEL$ removed;
the spectra plotted in red are the original unshifted spectra. Lines
are labeled assuming $\gamma_1$ of 110 $\VEL$.  A shift of 100
$\VEL$ corresponds to 0.37 \AA\ at these wavelengths.
\label{fig_gamma_vel}}

% fig 12
\figcaption[fuse_massrad.ps]{Constraints on the mass and radius of
the WD in U Gem for a K$_1$ velocity of 120 $\VEL$ and K$_2$
velocity of 300 $\VEL$. Constraints imposed for various values of
the gravitational redshift are shown; the black lines are the values
and errors derived by \cite{naylor2005_ugem}, using the
\cite{long1999} value for $\gamma_1$ and their value of $\gamma_2$.
The grey lines are similar, but use the average$\gamma_1$ from the
{\it FUSE} analysis. The solid black curve, labeled H-S, is the
Hamada-Salpeter mass radius relationship. The vertical dashed curves
indicate the mass for various inclinations. The horizontal
long-dashed lines is the range of photometric radii allowed by
single temperature, variable-z fits to Obs.\ 2. and the astrometric
distance to U Gem \label{fig_massrad}}

\pagebreak \clearpage \pagestyle{empty}

% tab1

%\input{tab_oblog}
%\begin{center}
\begin{deluxetable}{lcrcccc}
\setlength{\tabcolsep}{0.02in}
\tablecaption{Observation Log }
\tablehead{\colhead{Obs\#} &
 \colhead{FUSE~ID} &
 \colhead{Date} &
 \colhead{Start~(UT)} &
 \colhead{End~(UT)} &
 \colhead{Exptime~(s)} &
 \colhead{Days~since~peak}
}
\scriptsize
\tablewidth{0pt}\startdata
1...... &  A126 &  17~Mar~2000 &  11:43:20 &  20:34:16 &  12975 &  $\sim$10 \\
2...... &  P154 &  22-23~Feb~2001 &  17:35:19 &  09:08:52 &  13000 &  $\sim$135 \\
\enddata
\label{table_obs}
\end{deluxetable}
%\end{center}

%

% This table has been eliminated
% tab2
%\input{tab_lines}
%\input{tab2}
%

% tab3
%\input{table_norm}
%\begin{center}
\begin{deluxetable}{ccccccc}
\tablecaption{Uniform Temperature Solar-abundance WD Fits }
\tablehead{\colhead{Obs.} &
 \colhead{log(g)} &
 \colhead{Norm} &
 \colhead{R} &
 \colhead{T} &
 \colhead{$v\:\sin{(i)}$} &
 \colhead{$\CHINU$}
\\
\colhead{~} &
 \colhead{~} &
 \colhead{($10^{-23}$)} &
 \colhead{($10^{8}\:cm$)} &
  \colhead{(K)} &
 \colhead{($\VEL$)} &
 \colhead{~}
}
\scriptsize
\tablewidth{0pt}\startdata
1 &  8.0 &  3.7 &  5.3 &  40.7 &  163 &  7.2 \\
1 &  8.5 &  3.3 &  5.0 &  43.6 &  152 &  6.7 \\
1 &  9.0 &  3.0 &  4.8 &  47.1 &  135 &  6.5 \\
2 &  8.0 &  4.3 &  5.7 &  28.6 &  ~96 &  7.0 \\
2 &  8.5 &  3.4 &  5.1 &  30.3 &  ~90 &  6.8 \\
2 &  9.0 &  3.1 &  4.8 &  31.6 &  ~83 &  7.0 \\
\enddata
\label{tab_norm}
\end{deluxetable}
%\end{center}

% tab4
%\input{table_1wd}
%\begin{center}
\begin{deluxetable}{cccccccc}
\tablecaption{Uniform Temperature Variable Z WD Fits }
\tablehead{\colhead{Obs.} &
 \colhead{log(g)} &
 \colhead{Norm} &
 \colhead{R} &
 \colhead{T} &
 \colhead{z} &
 \colhead{$v\:\sin{(i)}$} &
 \colhead{$\CHINU$}
\\
\colhead{~} &
 \colhead{~} &
 \colhead{($10^{-23}$)} &
 \colhead{($10^{8}\:cm$)} &
 \colhead{(1000 K)} &
 \colhead{~} &
 \colhead{($\VEL$)} &
 \colhead{~}
}
\scriptsize
\tablewidth{0pt}\startdata
1 &  8.0 &  4.0 &  5.3 &  40.6 &  1.3 &  170 &  7.1 \\
1 &  8.5 &  3.3 &  5.1 &  43.5 &  1.4 &  160 &  6.7 \\
1 &  9.0 &  3.0 &  4.8 &  46.8 &  1.5 &  148 &  6.4 \\
2 &  8.0 &  5.1 &  6.2 &  28.3 &  3.3 &  140 &  6.3 \\
2 &  8.5 &  4.3 &  5.7 &  29.7 &  3.9 &  130 &  6.0 \\
2 &  9.0 &  3.9 &  5.5 &  30.9 &  4.7 &  116 &  5.9 \\
\enddata
\label{tab_1wd}
\end{deluxetable}
%\end{center}

%

% tab5
%\input{table_2wd}
%\begin{center}
\begin{deluxetable}{ccccccccccc}
%\rotate
\setlength{\tabcolsep}{0.02in}
\tablecaption{Multi-Temperature WD Fits } \tablehead{\colhead{Obs.}
&
 \colhead{log(g)} &
 \colhead{Norm$_1$} &
 \colhead{T$_1$} &
 \colhead{z$_1$} &
 \colhead{$v\:\sin{(i)}_1$} &
 \colhead{Norm$_2$} &
 \colhead{T$_2$} &
 \colhead{z$_2$} &
 \colhead{$v\:\sin{(i)}_2$} &
 \colhead{$\CHINU$}
\\
\colhead{~} &
 \colhead{~} &
 \colhead{($10^{-23}$)} &
 \colhead{(1000 K)} &
 \colhead{~} &
 \colhead{($\VEL$)} &
 \colhead{($10^{-23}$)} &
 \colhead{(1000 K)} &
 \colhead{~} &
 \colhead{($\VEL$)} &
 \colhead{~}
}
\scriptsize
\tablewidth{0pt}\startdata
1 &  8.0 &  6.5 &  26.1 &  ~1.5 &  133 &  1.1 &  65.8 & ~9.1 &  236 &  5.5 \\
1 &  8.5 &  4.5 &  28.5 &  ~1.5 &  ~87 &  1.0 &  70.0 & ~8.9 &  243 &  5.6 \\
1 &  9.0 &  3.0 &  31.0 &  10.0 &  186 &  1.4 &  60.7 & ~0.8 &  ~79 &  5.7 \\
2 &  8.0 &  7.0 &  25.0 &  ~4.3 &  150 &  0.6 &  38.1 & ~3.8 &  ~65 &  5.6 \\
2 &  8.5 &  4.6 &  26.1 &  ~5.7 &  159 &  1.1 &  34.5 & ~3.4 &  ~76 &  5.7 \\
2 &  9.0 &  3.2 &  28.8 &  10.0 &  139 &  1.2 &  33.9 & ~1.7 &  ~71 &  5.8 \\
\enddata
\label{tab_2wd}
\end{deluxetable}
%\end{center}

%

% tab 6
%\input{table_veil}
%\begin{center}
\begin{deluxetable}{cccccccc}
\setlength{\tabcolsep}{0.02in}
\tablecaption{WD \& Absorbing Screen Fits of Obs.\ 2 } \tablehead{
 \colhead{Spectrum} &
 \colhead{Norm} &
 \colhead{T$_{WD}$} &
 \colhead{$v\:\sin{(i)}$} &
 \colhead{log~(NH)} &
 \colhead{T$_{abs}$} &
 \colhead{$v_{abs}$} &
 \colhead{$\CHINU^{~a}$}
\\
\colhead{~} &
 \colhead{($10^{-23}$)} &
 \colhead{(1000~K)} &
 \colhead{($\VEL$)} &
 \colhead{(cm$^{-2}$)} &
 \colhead{(1000~K)} &
 \colhead{($\VEL$)} &
 \colhead{}
}
\scriptsize
\tablewidth{0pt}\startdata
Unabsorbed &  3.6 &  30.4 &  54 &  19.6 &  16.8 &  170 &  3.4 \\
Phase~0.2-0.35 &  4.5 &  29.4 &  96 &  20.7 &  10.3 &  250 &  3.4 \\
Phase~0.6-0.85 &  3.3 &  29.9 &  55 &  21.3 &  12.0 &  160 &  3.1 \\
\tablenotetext{a}{ See text for discussion of \CHINU\ in these
fits.}
\enddata
\label{tab_veil}
\end{deluxetable}
%\end{center}

% tab 7
%\input{tab_gamma}
%\begin{center}
\begin{deluxetable}{rlcc}
\tablecaption{$\gamma$ Velocities of Selected Lines }
\tablehead{\colhead{~} & 
 \colhead{Laboratory} & 
 \colhead{Obs.~1} & 
 \colhead{Obs.~2} 
\\
\colhead{Transition} & 
 \colhead{Wavelength} & 
 \colhead{$\gamma$} & 
 \colhead{$\gamma$} 
\\
\colhead{~} & 
 \colhead{(\AA)} & 
 \colhead{($\VEL$)} & 
 \colhead{($\VEL$)} 
}
\scriptsize
\tablewidth{0pt}\startdata
\ion{S}{4}$\lambda$1062 &  1062.662 &  167 &  135 \\ 
\ion{Si}{4}$\lambda$1067 &  1066.6498 &  151 &  125 \\ 
\ion{S}{4}$\lambda$1073 &  1072.974 &  147 &  123 \\ 
\ion{Si}{3}$\lambda$1108 &  1108.3579 &  149 &  152 \\ 
\ion{Si}{3}$\lambda$1110 &  1109.9696 &  118 &  132 \\ 
\ion{Si}{3}$\lambda$1113 &  1113.2296 &  136 &  119 \\ 
\ion{Si}{4}$\lambda$1122 &  1122.4849 &  146 &  132 \\ 
\enddata 
\label{tab_gamma}
\end{deluxetable}
%\end{center}

\pagestyle{empty}
% fig 1
\begin{figure}
%\plotone{aavso.ps}
\plotone{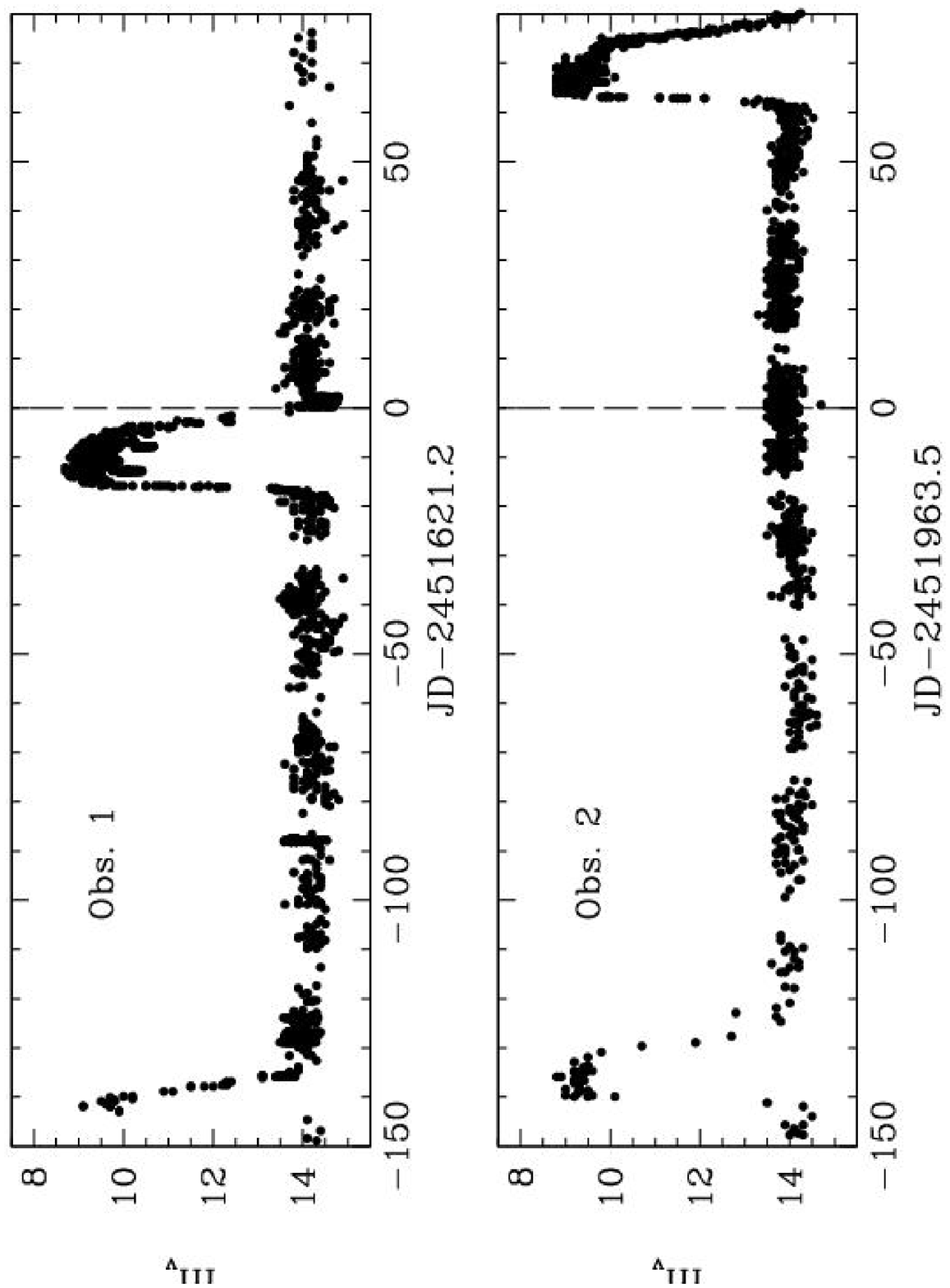}
\end{figure}

% fig 2
%\begin{figure}
%\plotone{shape.ps}
%\plotone{f02.eps}
%\end{figure}

% fig 3
\begin{figure}
%\plotone{lines.ps}
\plotone{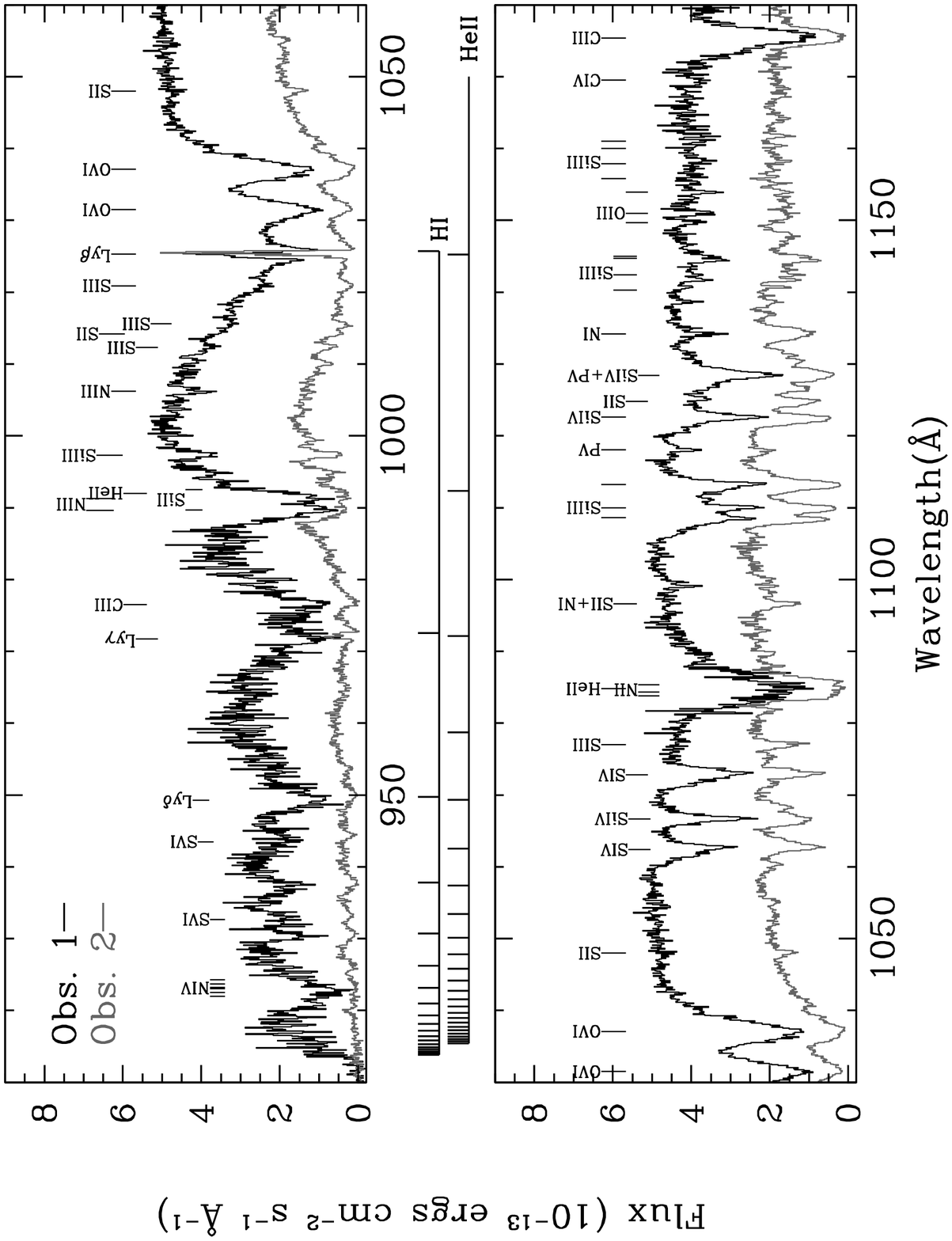}
\end{figure}

% fig 4
\begin{figure}
%\plotone{dips.ps}
\plotone{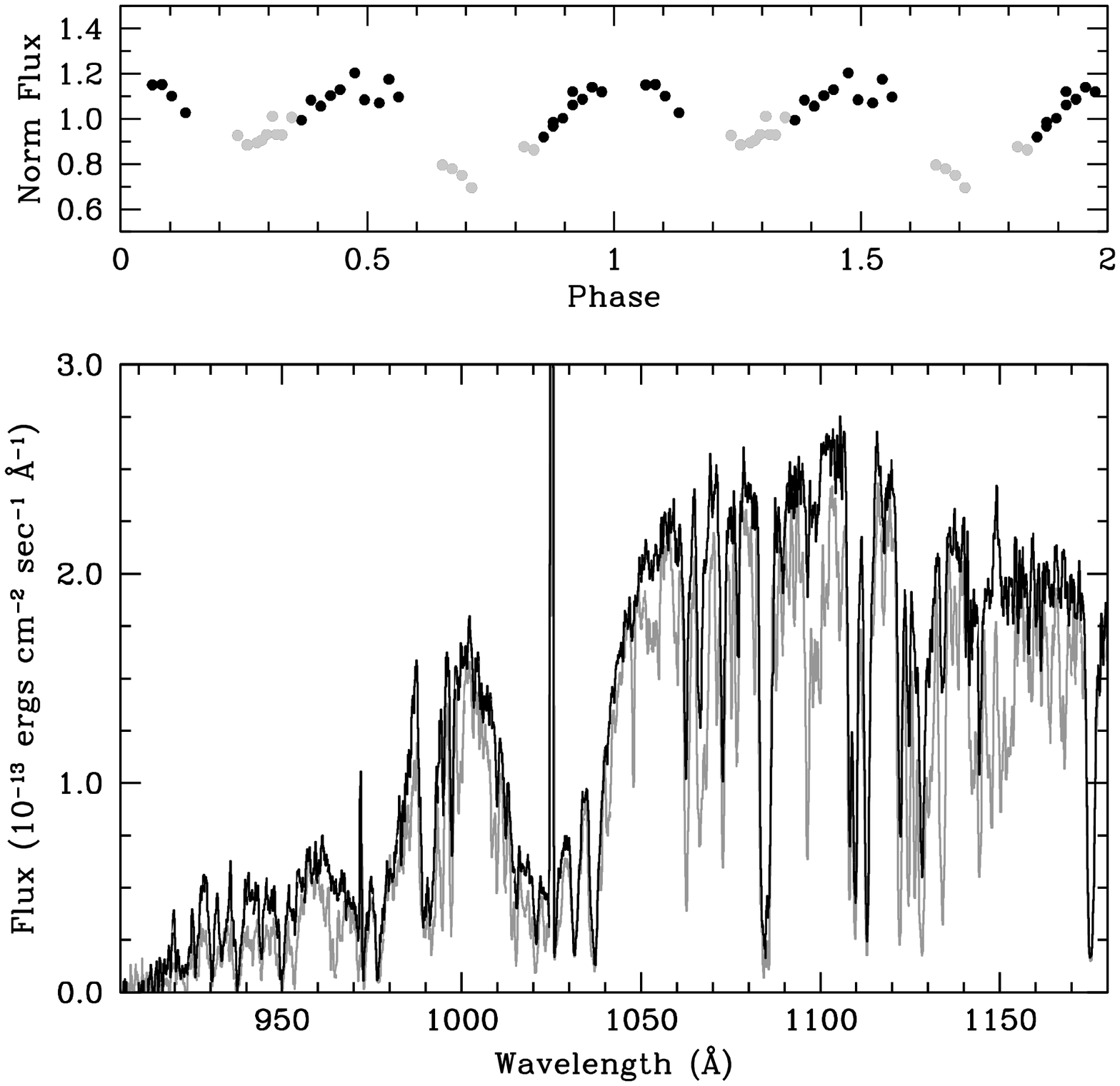}
\end{figure}

% fig 5
% Normal abundance fit to Obs 1
\begin{figure}
%\plotone{obs1_norm85_0511.ps}
\plotone{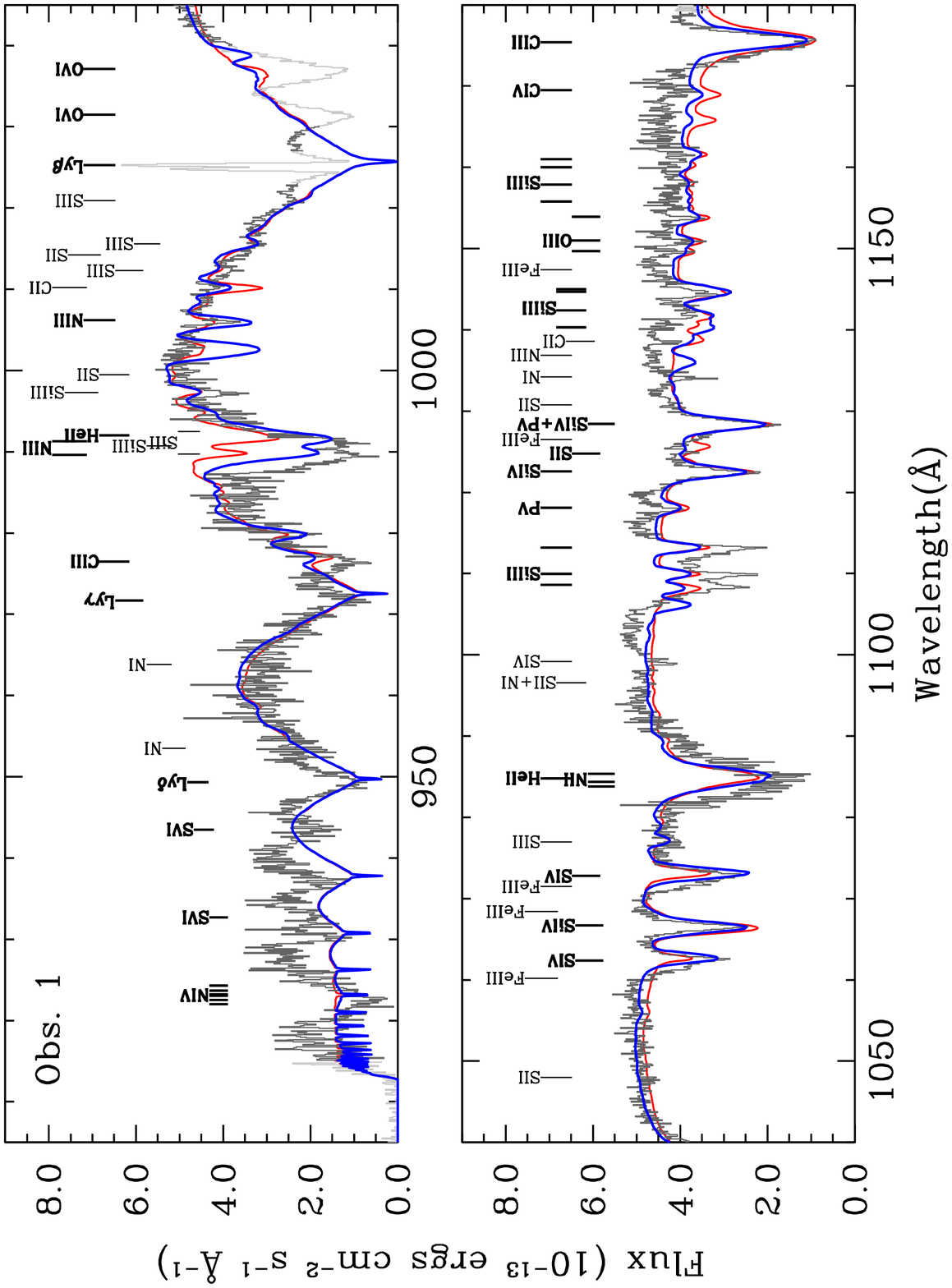}
\end{figure}

\pagebreak

% fig 6
% zscaled abundance fit to Obs2
\begin{figure}
%\plotone{ob2_wdg85_0511.ps}
\plotone{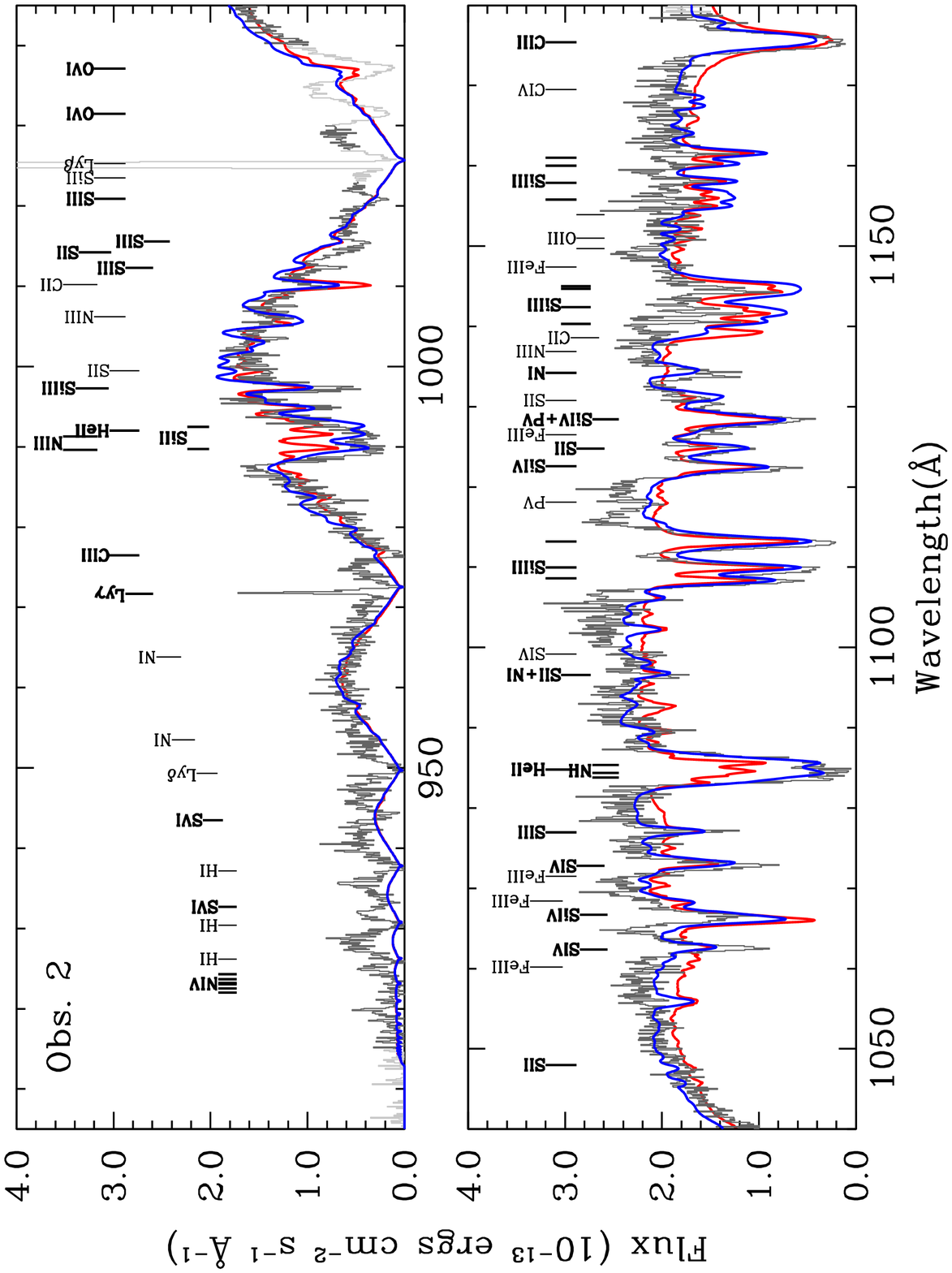}
\end{figure}

% fig 7
% 2 WD model for observation 1 with g85
\begin{figure}
%\plotone{obs1_2wd85_0511.ps}
\plotone{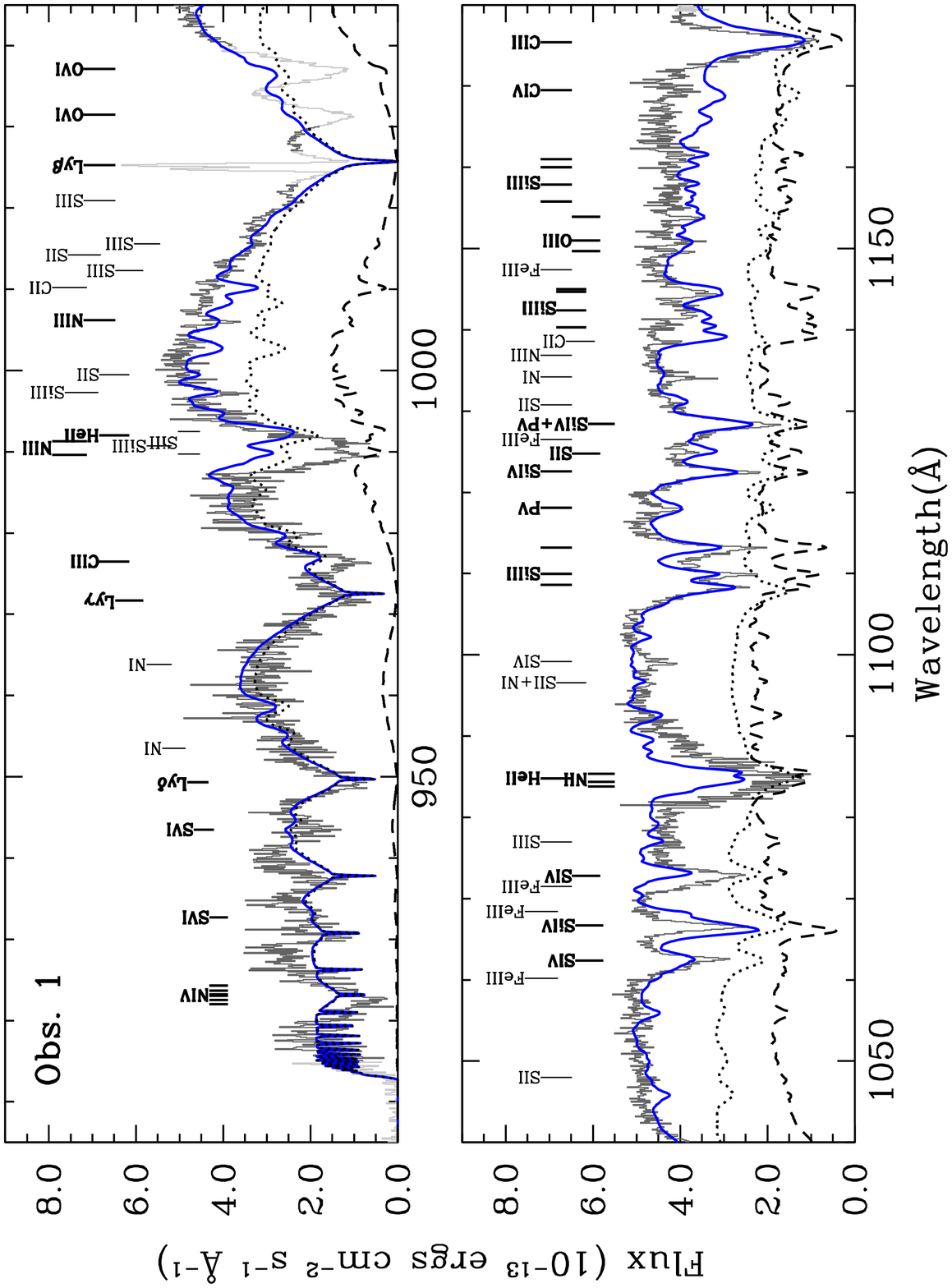}
\end{figure}

% New figure

\begin{figure}
\plotone{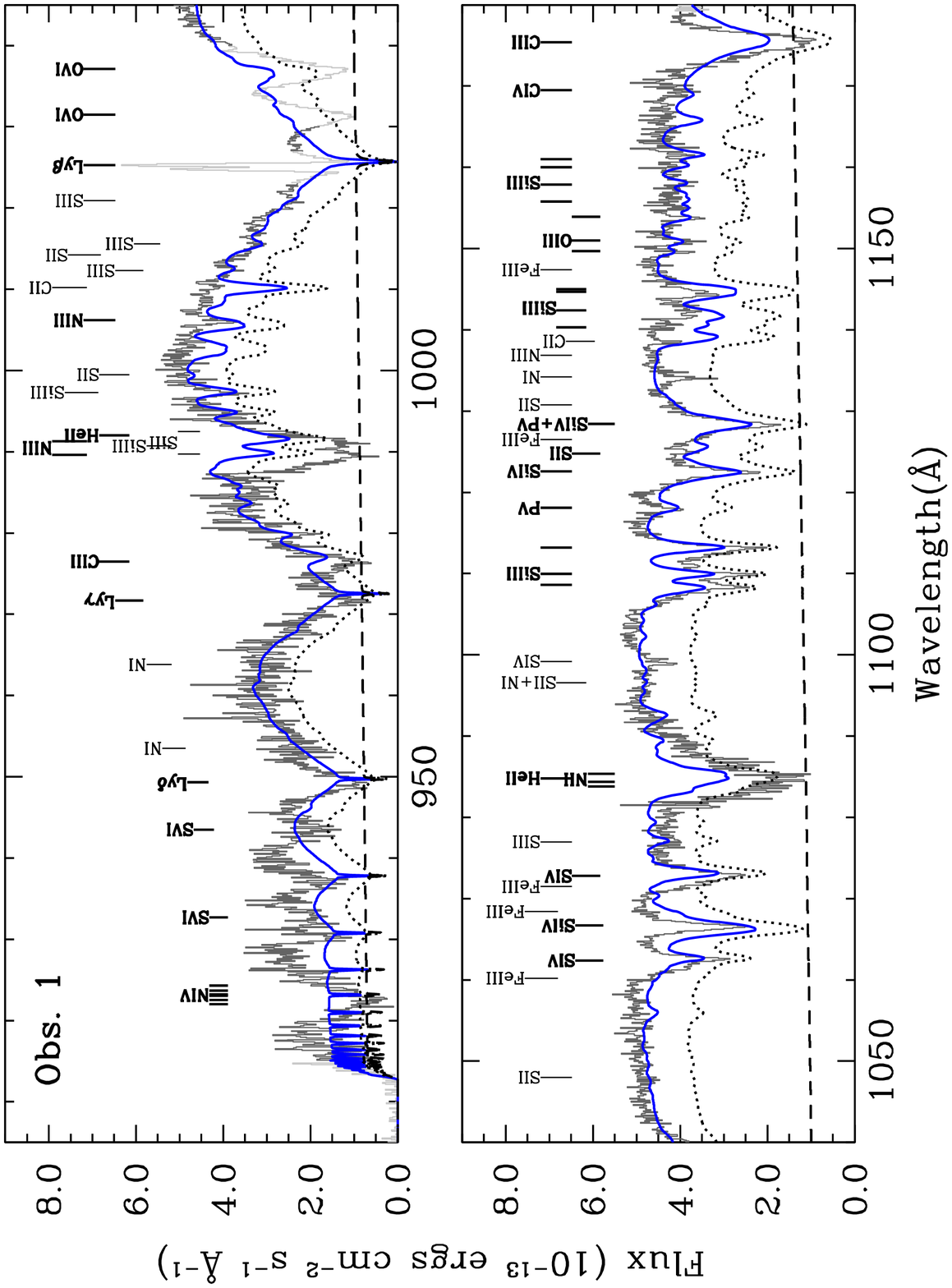}
\end{figure}

% fig 8
\begin{figure}
%\plotone{abs_1125.ps} %\plotone{many_abs_fit.ps}
\plotone{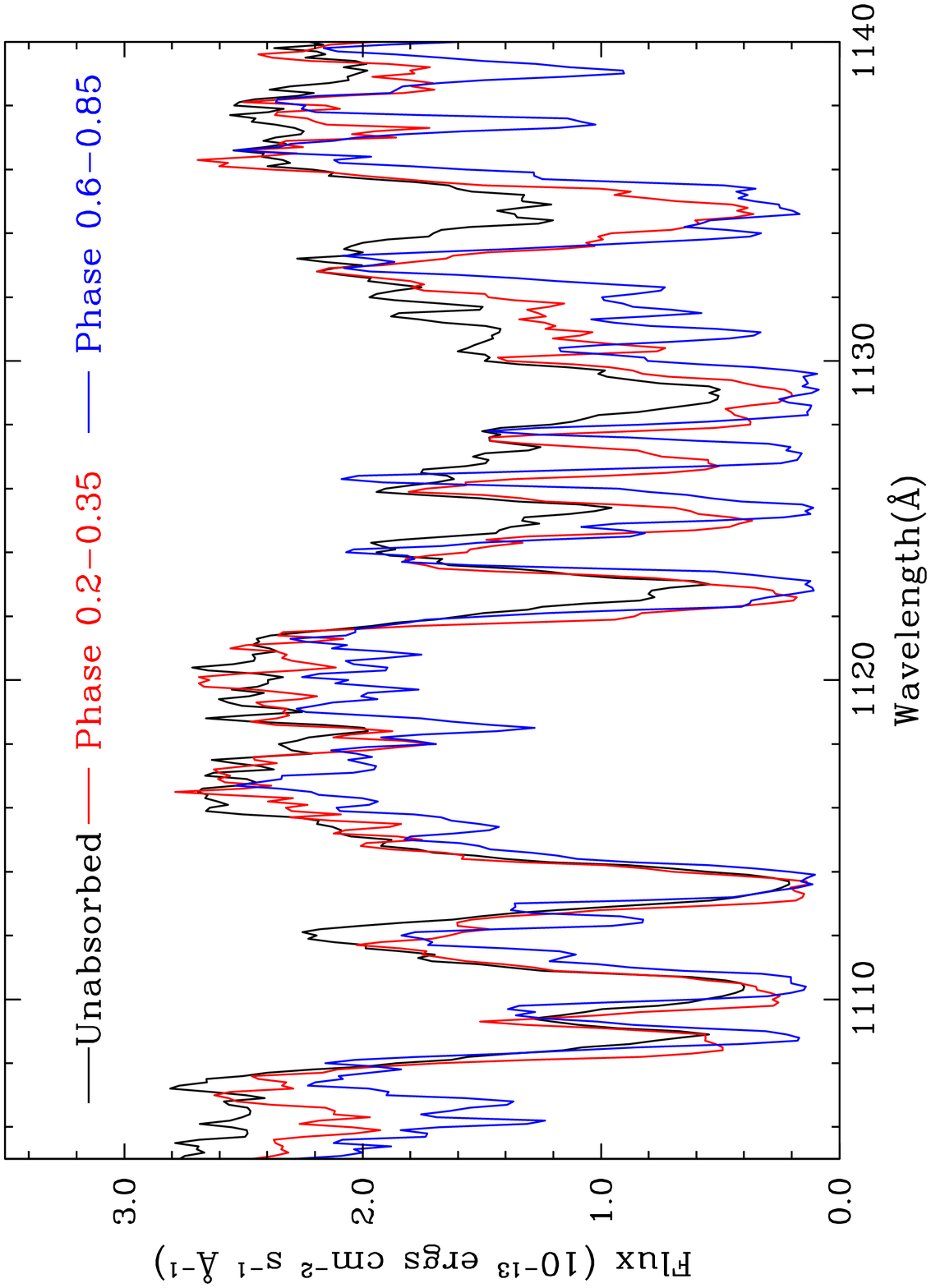}
\end{figure}

% fig 9
\begin{figure}
%\plotone{veils.ps}
\plotone{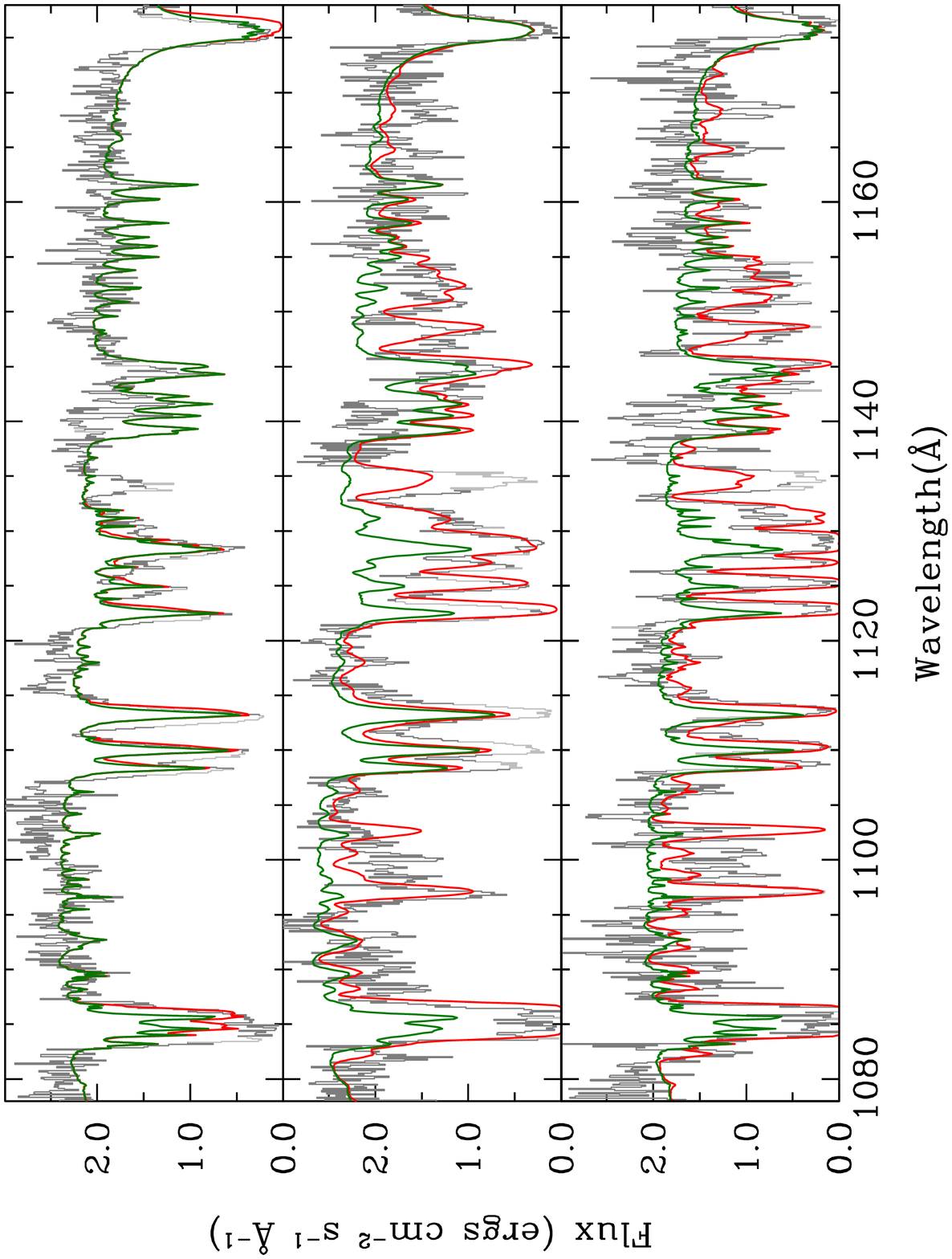}
\end{figure}

% fig 10
\begin{figure}
%\plotone{K1.ps}
\plotone{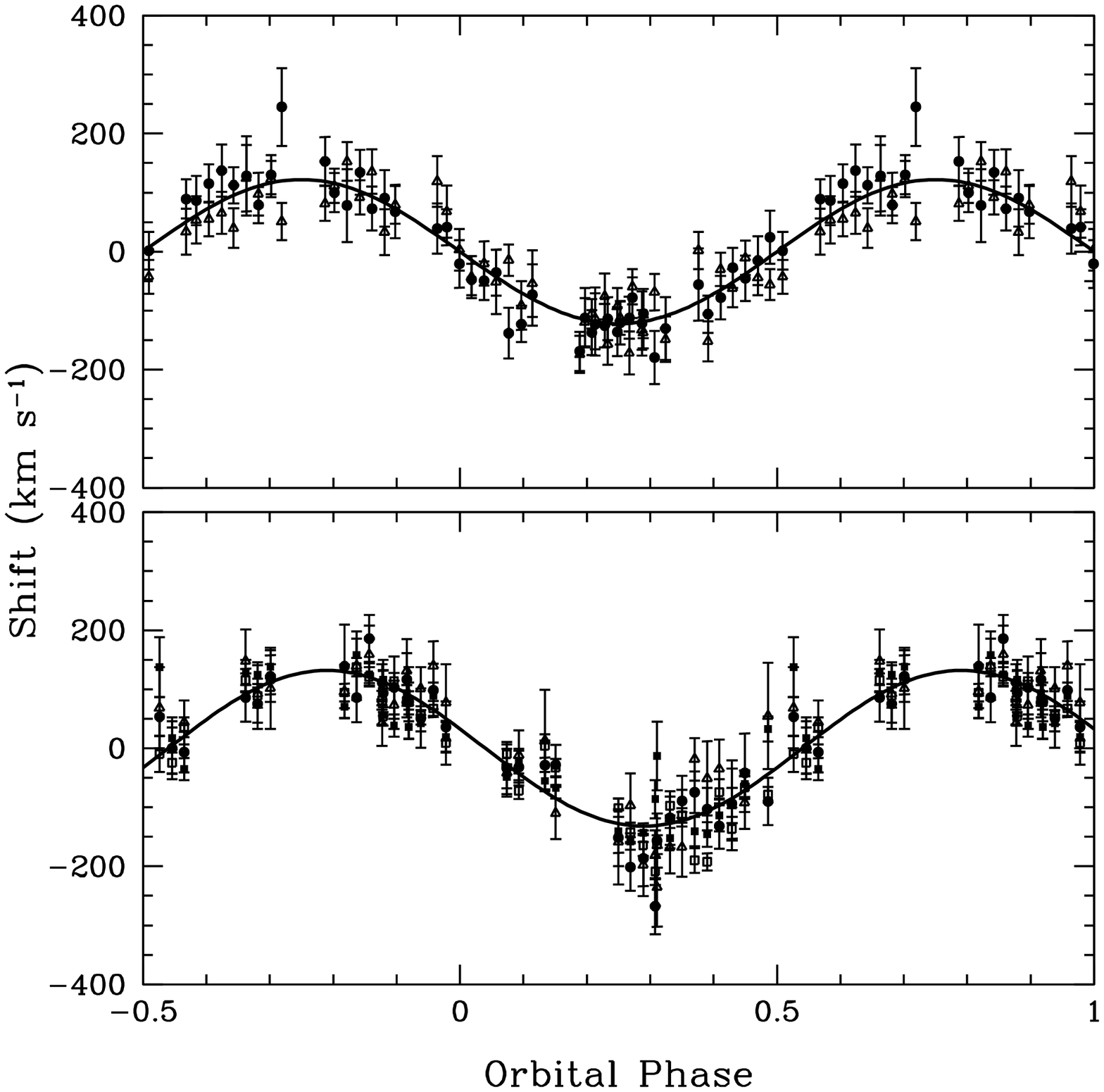}
\end{figure}

% fig 11
\begin{figure}
%\plotone{fig_vshift.eps}
\plotone{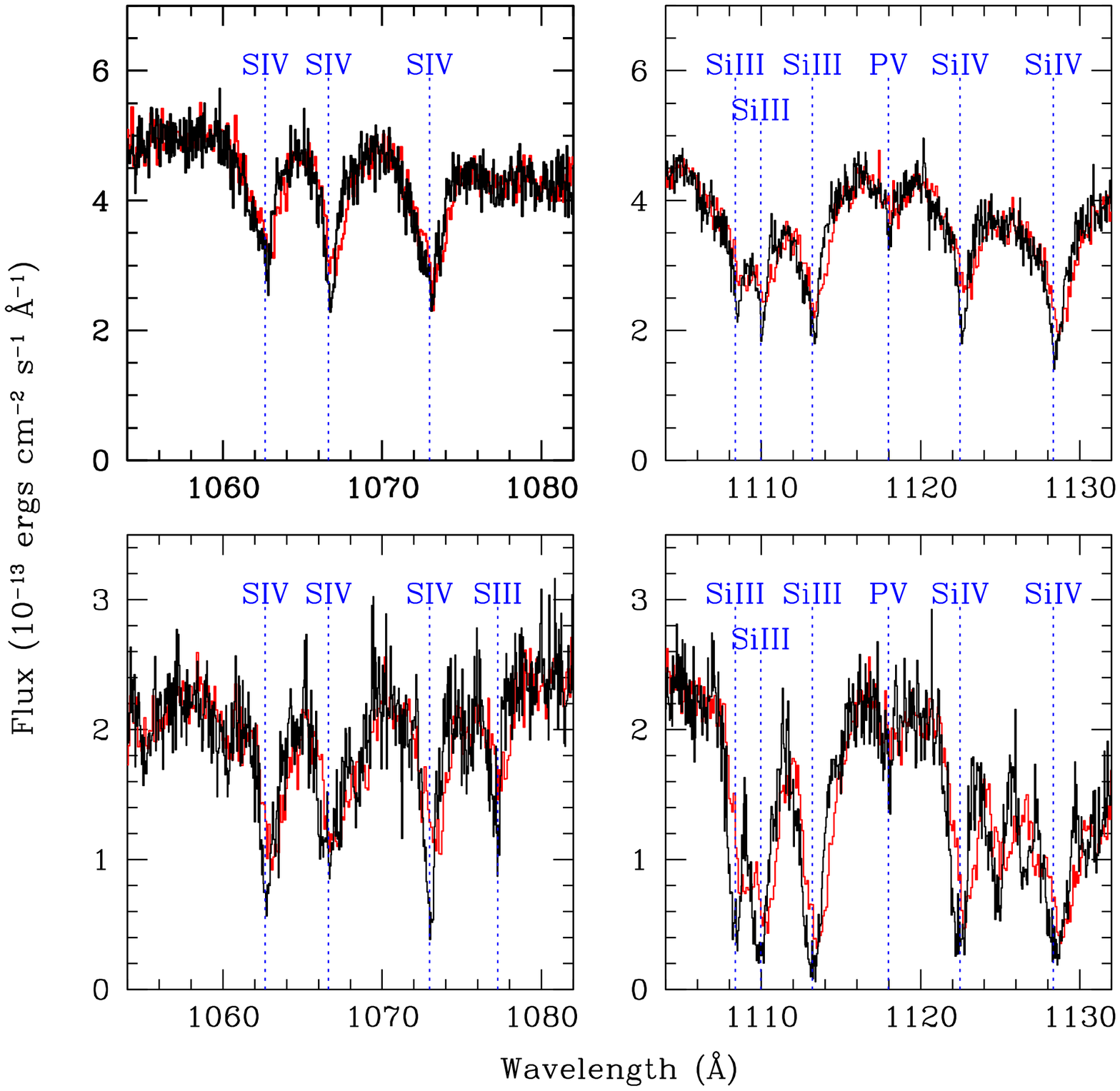}
\end{figure}

\clearpage

% fig 12
\begin{figure}
%\plotone{fuse_massrad.ps}
\plotone{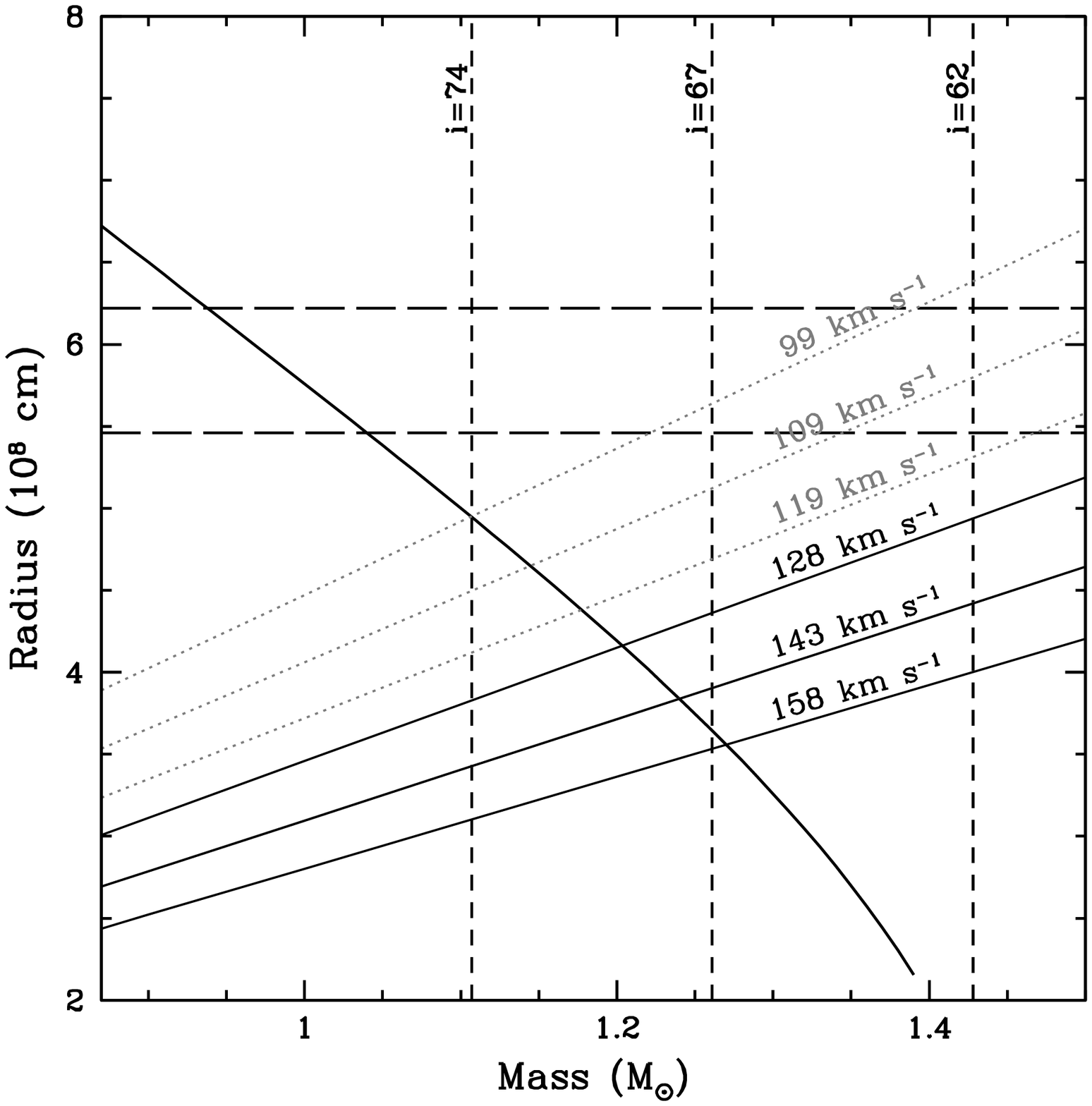}
\end{figure}

\end{document}